\PassOptionsToPackage{table}{xcolor}
\documentclass{article}
\usepackage{iclr2027_conference,times}

\usepackage{amsmath}
\usepackage{booktabs}
\usepackage{multirow}
\usepackage{tabularx}
\usepackage[table]{xcolor}
\usepackage{tcolorbox}
\usepackage{enumitem}
\usepackage{graphicx}
\usepackage{capt-of}
\usepackage{afterpage}
\usepackage{placeins}
\usepackage{hyperref}
\usepackage{microtype}
\usepackage{url}
\usepackage[scaled=0.95]{helvet}

\title{WideSWE: Can Coding Agents Coordinate Changes Across Repositories?}

\author{%
Baoyi Wang\textsuperscript{1,*}\quad
Xingliang Wang\textsuperscript{1,*}\quad
Jinyang Wu\textsuperscript{2}\quad
Keming Wu\textsuperscript{2}\\
Chen Zhi\textsuperscript{1,\textdagger}\quad
Jianwei Yin\textsuperscript{1}
}

\iclrfinalcopy

\newcommand{\benchmark}{\textsc{WideSWE}}
\newcommand{\fpt}{F2P}
\newcommand{\ppt}{P2P}
\newcommand{\system}{Codex CLI--GPT-5.6-sol}
\definecolor{rqheader}{HTML}{EEF0F2}
\definecolor{rqcase}{HTML}{EDF4FA}
\definecolor{rqgroup}{HTML}{304B5B}
\definecolor{rqup}{HTML}{28785C}
\definecolor{rqdown}{HTML}{B45A1B}
\definecolor{rqcodex}{HTML}{285F8F}
\definecolor{rqcc}{HTML}{76518C}

\hypersetup{
    pdftitle={WideSWE: Can Coding Agents Coordinate Changes Across Repositories?},
    pdfauthor={Baoyi Wang, Xingliang Wang, Jinyang Wu, Keming Wu, Chen Zhi, Jianwei Yin},
    colorlinks=true,
    citecolor=rqcodex,
    linkcolor=rqcodex,
    urlcolor=rqcodex
}

\newtcolorbox{requirementquote}{
    colback=rqcase!45!white,
    colframe=rqcodex!30!white,
    boxrule=0.5pt,
    arc=2pt,
    boxsep=0pt,
    left=10pt, right=10pt, top=6pt, bottom=6pt,
    before skip=6pt, after skip=6pt
}

\definecolor{arxivpanel}{HTML}{F7F8FB}
\makeatletter
\newcommand{\arxivtitle}{%
    {\sffamily\bfseries\fontsize{19}{21.5}\selectfont\raggedright\@title\par}%
    \vspace{10pt}%
    {\sffamily\bfseries\fontsize{10}{13}\selectfont\raggedright\@author\par}%
}
\makeatother

\begin{document}
\fancyhead{}
\renewcommand{\headrulewidth}{0pt}

\begin{tcolorbox}[
    colback=arxivpanel,
    colframe=arxivpanel,
    boxrule=0pt,
    arc=10pt,
    boxsep=0pt,
    left=16pt, right=16pt, top=17pt, bottom=17pt,
    before skip=0pt, after skip=18pt
]
\arxivtitle
\vspace{8pt}
{\small\noindent
\textsuperscript{1}Zhejiang University, China\quad
\textsuperscript{2}Tsinghua University, China\par
\textsuperscript{*}Both authors contributed equally to this research.\par
\textsuperscript{\textdagger}Corresponding author: Chen Zhi.\par}
\vspace{14pt}
\begingroup
\renewenvironment{abstract}{\par\noindent}{\par}
\begin{abstract}
Coding-agent evaluation has progressed from resolving individual issues to carrying out long-horizon development, yet task completion is still largely assessed within a single codebase. In software ecosystems, many features and bug fixes require coordinated changes across multiple repositories. We introduce \benchmark{} to evaluate coding agents on such cross-repository tasks. Mining and reviewing changes across 103 software ecosystems yields 120 real-world tasks, balanced between 60 bug fixes and 60 features. We derive prompts from related issues and pull requests. We systematically review and adapt hidden tests to support diverse correct implementations while preserving required behavior and regression checks. Across seven agent configurations, full task success ranges from 10.83\% to 42.50\%, with the configuration pairing Codex CLI with GPT-5.6-sol achieving the highest rate. Trajectories show agents failing to identify necessary changes, recognizing changes but leaving them unfinished, or modifying the required repositories without fully satisfying the request. To examine whether working on one repository at a time can alleviate these difficulties, we compare it with joint execution under identical prompts. Independent execution mainly recovers omitted work and is less effective at correcting previously attempted but unsuccessful implementations. Joint execution can use information from related repositories to guide implementation and verification.
\end{abstract}

\endgroup
\vspace{13pt}
\noindent\href{https://github.com/ZJU-ACES-ISE/WideSWE}{\raisebox{-2pt}{\includegraphics[height=11pt]{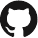}}}\hspace{4pt}{\sffamily\bfseries Code:}\enspace
\url{https://github.com/ZJU-ACES-ISE/WideSWE}
\end{tcolorbox}
\begingroup
\renewcommand{\thefootnote}{\fnsymbol{footnote}}
\footnotetext[2]{Emails: \texttt{\{wangbaoyi,wangxingliang,zjuzhichen\}@zju.edu.cn},\newline
\hspace*{1.8em}\texttt{\{wu-jy23,wukm25\}@mails.tsinghua.edu.cn}, \texttt{zjuyjw@cs.zju.edu.cn}.}
\endgroup

\section{Introduction}
\label{sec:introduction}

Large language models (LLMs) now serve as the foundation of coding agents that can inspect repositories, edit multiple files, run tests, and iteratively carry out software-engineering tasks~\citep{liu2024repoqa,wang2025inspectcoder}. Evaluation has expanded beyond isolated code generation~\citep{chen2021evaluating,austin2021program,wang2025exploracoder,zhang2026code2bench} toward autonomous work in real repositories~\citep{pan2024training}. SWE-bench asks agents to resolve real GitHub issues in a repository~\citep{jimenez2024swe}. More recent benchmarks extend this scope: DeepSWE evaluates original, long-horizon engineering tasks~\citep{huang2026deepswe}, while ProgramBench requires rebuilding complete programs from reference executables and documentation~\citep{yang2026programbench}. Despite this broader scope, these benchmarks still focus primarily on completing tasks in a single repository.

In software ecosystems, however, a single feature or bug fix may require coordinated changes across several repositories~\citep{blincoe2019reference,ma2017developers}. For example, introducing a shared feature across Sentry's language SDKs\footnote{\url{https://github.com/getsentry}} requires separate implementations in its Go, Python, and Ruby repositories. Other changes involve dependencies: a new capability in Sentry's PHP SDK also requires corresponding updates to two other repositories that depend on it. Cross-repository links are common in practice: our analysis of 1,729,171 PRs across 103 software ecosystems identified 109,233 PR records explicitly referencing another repository in the same ecosystem. Can coding agents handle this coordination and complete the request across repositories?

\begin{figure}[t]
    \centering
    \includegraphics[width=\linewidth]{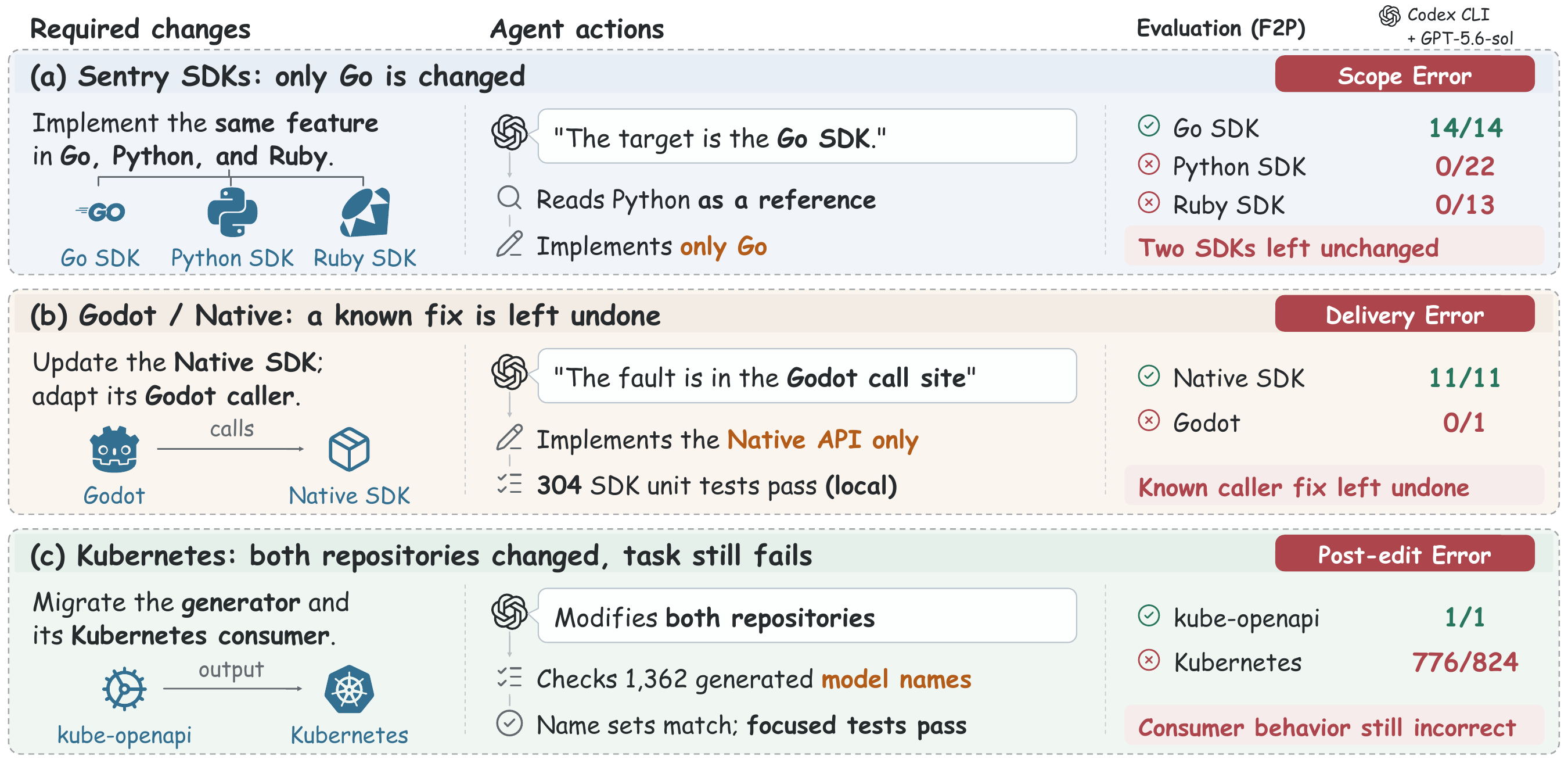}
    \caption{Three observed cross-repository failures in \system{} runs.}
    \label{fig:coordination-challenges}
\end{figure}

To examine this question, we first evaluate an agent configuration pairing Codex CLI with GPT-5.6-sol. As Figure~\ref{fig:coordination-challenges} illustrates, failures occurred in identifying the full scope of affected repositories, delivering the required changes in each, and satisfying task requirements after editing. In Sentry, the agent narrows a three-SDK requirement to Go, leaving Python and Ruby unchanged. In the Godot/Native task, both repositories require changes; the agent recognizes this scope but modifies only the Native SDK. In Kubernetes, it identifies and modifies both required repositories, but the changes are incorrect and the task remains unsolved.

To assess agents' ability to handle these challenges, we build \benchmark{}, grouping changes that implement the same feature or fix across repositories into a single task and evaluating the request as a whole. From the 200 GitHub organizations ranked highest by total repository stars,\footnote{Gitstar Ranking: \url{https://gitstar-ranking.com/organizations}, collected on June 5, 2026.} we identify 103 active software ecosystems whose repositories support a common product, platform, or technology. We mine linked, merged PRs and use manual review and executable validation to identify tasks requiring substantive changes in multiple repositories. Balancing task types yields 120 cases: 60 bug fixes and 60 features. We merge requirements from the original issues and PRs into task prompts. We also address mismatches between PR requirements and inherited tests, preserving required behavior while supporting alternative correct implementations.

Our experiments show that agents do not reliably carry a shared requirement through all affected repositories. \system{} solves only 42.50\% of cases. In 48 of its 69 failed cases, at least one repository passes all its fail-to-pass (\fpt{}) tests while another does not. The trajectories show the agent missing part of the required scope or stopping after tests pass in one repository, leaving already identified downstream work unfinished. We further compare joint execution in the ecosystem workspace with independent execution in each target repository. Across 89 cases with identical prompts, joint execution solves 36 and independent execution solves 32, with 20 outcome reversals. Trajectory analysis reveals two contrasting patterns: independent execution can recover omitted work by narrowing the scope, but can also lose behavioral references from related repositories.

Our contributions are threefold:
\begin{itemize}[leftmargin=*,nosep]
    \item \textbf{A task formulation and benchmark for cross-repository coding.} We formulate cross-repository task completion as implementing one shared feature or bug fix across multiple target repositories. We introduce \benchmark{}, comprising 120 real-world tasks across 41 software ecosystems.
    \item \textbf{Requirement-aligned test review.} We identify mismatches between task requirements and inherited tests, and address them through rule-guided manual revisions that preserve required functionality while supporting alternative correct implementations.
    \item \textbf{An empirical study of cross-repository task completion.} We evaluate seven agent configurations, examining differences across models, scaffolds, task types, repository counts, and language diversity. We combine joint-versus-independent comparisons with trajectory analysis to provide empirical insights into cross-repository task completion.
\end{itemize}

\section{The \benchmark{} Benchmark}
\label{sec:benchmark}

\begin{figure}[!t]
    \centering
    \includegraphics[width=\linewidth]{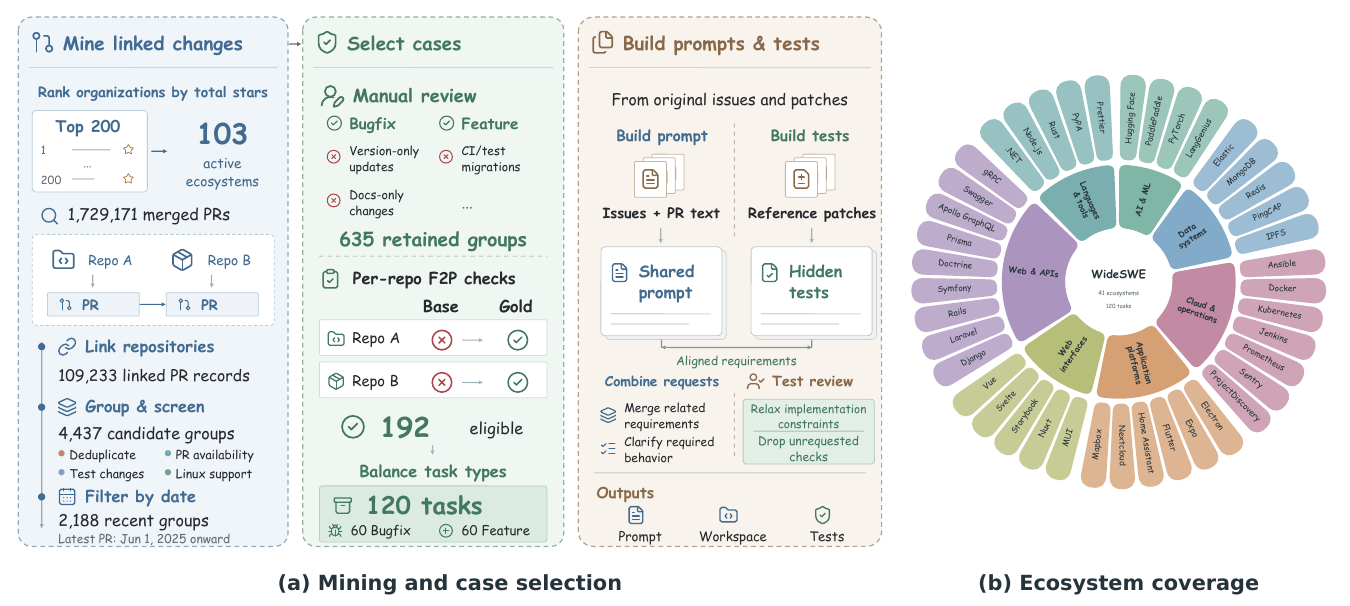}
    \caption{\benchmark{} construction and coverage. (a) Mining and case selection. (b) The 41 ecosystems grouped by application area.}
    \label{fig:benchmark-selection}
    \label{fig:ecosystem-map}
\end{figure}

\subsection{Task Formulation}
\label{sec:task-formulation}

Each \benchmark{} task represents one feature or bug fix requiring coordinated changes in at least two \emph{target repositories}. Coordination involves not only handling explicit interface dependencies but also autonomously identifying which repositories require changes to fulfill the shared request and completing those changes while preserving existing behavior and cross-repository consistency.

For task $c$, let $p_c$ denote the request and $\mathcal{W}_c$ the historical workspace containing repositories $\mathcal{R}_c$. The target set $\mathcal{T}_c \subseteq \mathcal{R}_c$, with $|\mathcal{T}_c| \geq 2$, comprises the repositories that require substantive changes to fulfill the request. Repositories in $\mathcal{R}_c \setminus \mathcal{T}_c$ provide additional ecosystem context. In addition to the targets, agents can access up to 20 context repositories in a task workspace (Figure~\ref{fig:dataset-statistics} in Appendix~\ref{app:dataset-statistics}).

Given $p_c$ and $\mathcal{W}_c$, an agent $A$ can inspect and modify the workspace and produces a collection of repository patches $\Delta_c$:
\begin{equation}
    \Delta_c = A(p_c, \mathcal{W}_c), \qquad
    \mathcal{W}'_c = \mathcal{W}_c \oplus \Delta_c,
    \label{eq:task-formulation}
\end{equation}
where $\oplus$ applies the patches to the original snapshots and $\mathcal{W}'_c$ is the resulting workspace. Target repositories determine the scoring scope; context repositories remain available for inspection and modification. Reference patches and hidden tests are withheld from the agent.

\begin{table}[!t]
\caption{Main results by task type. Codex denotes Codex CLI, and CC denotes Claude Code. Model headers abbreviate GPT-5.6-sol, Claude Opus 5, Gemini 3.8 Flash, DeepSeek V4 Pro, and Qwen 3.8 Max. Case/Repo denote task/repository success (\%); F2P/P2P denote F2P-complete/P2P-preserved repositories (\%). API denotes mean API requests per task. Bold marks the highest rate in each row.}
\label{tab:main-results}
\centering
\small
\setlength{\tabcolsep}{2pt}
\renewcommand{\arraystretch}{1.12}
\begin{tabularx}{\linewidth}{@{}ll*{7}{>{\centering\arraybackslash}X}@{}}
\toprule
\rowcolor{rqheader}
Task & Metric & \shortstack{{\footnotesize GPT-5.6}\\{\scriptsize \textcolor{rqcodex}{+ Codex}}} & \shortstack{{\footnotesize GPT-5.6}\\{\scriptsize \textcolor{rqcc}{+ CC}}} & \shortstack{{\footnotesize Opus 5}\\{\scriptsize \textcolor{rqcc}{+ CC}}} & \shortstack{{\footnotesize Gemini}\\{\scriptsize \textcolor{rqcc}{+ CC}}} & \shortstack{{\footnotesize DeepSeek}\\{\scriptsize \textcolor{rqcc}{+ CC}}} & \shortstack{{\footnotesize Qwen}\\{\scriptsize \textcolor{rqcc}{+ CC}}} & \shortstack{{\footnotesize GLM 5.3}\\{\scriptsize \textcolor{rqcc}{+ CC}}} \\
\midrule
\rowcolor{rqcase}
 & \textbf{Case}\,\textcolor{rqup}{$\uparrow$} & 41.67 & 33.33 & 41.67 & 21.67 & 38.33 & \textbf{48.33} & 30.00 \\
 & Repo\,\textcolor{rqup}{$\uparrow$} & 61.67 & 55.83 & 63.33 & 26.67 & 60.00 & \textbf{66.67} & 52.50 \\
 & \fpt{}\,\textcolor{rqup}{$\uparrow$} & 62.50 & 58.33 & 64.17 & 28.33 & 62.50 & \textbf{67.50} & 52.50 \\
 & \ppt{}\,\textcolor{rqup}{$\uparrow$} & 95.37 & 91.67 & \textbf{96.30} & \textbf{96.30} & 92.59 & \textbf{96.30} & 89.81 \\
\multirow{-5}{*}{\textcolor{rqgroup}{\textbf{Bugfix}}} & API\,\textcolor{rqdown}{$\downarrow$} & 62.8 & 76.9 & 118.7 & 178.4 & 117.3 & 141.5 & 113.9 \\
\midrule
\rowcolor{rqcase}
 & \textbf{Case}\,\textcolor{rqup}{$\uparrow$} & \textbf{43.33} & 31.67 & 28.33 & 0.00 & 15.00 & 26.67 & 10.00 \\
 & Repo\,\textcolor{rqup}{$\uparrow$} & \textbf{65.41} & 57.89 & 48.12 & 6.02 & 36.84 & 48.12 & 28.57 \\
 & \fpt{}\,\textcolor{rqup}{$\uparrow$} & \textbf{68.42} & 61.65 & 48.87 & 6.02 & 39.10 & 48.12 & 30.83 \\
 & \ppt{}\,\textcolor{rqup}{$\uparrow$} & 86.73 & 78.76 & 85.84 & \textbf{95.58} & 77.88 & 88.50 & 78.76 \\
\multirow{-5}{*}{\textcolor{rqgroup}{\textbf{Feature}}} & API\,\textcolor{rqdown}{$\downarrow$} & 110.7 & 191.4 & 201.8 & 113.9 & 203.7 & 212.7 & 188.2 \\
\midrule
\rowcolor{rqcase}
 & \textbf{Case}\,\textcolor{rqup}{$\uparrow$} & \textbf{42.50} & 32.50 & 35.00 & 10.83 & 26.67 & 37.50 & 20.00 \\
 & Repo\,\textcolor{rqup}{$\uparrow$} & \textbf{63.64} & 56.92 & 55.34 & 15.81 & 47.83 & 56.92 & 39.92 \\
 & \fpt{}\,\textcolor{rqup}{$\uparrow$} & \textbf{65.61} & 60.08 & 56.13 & 16.60 & 50.20 & 57.31 & 41.11 \\
 & \ppt{}\,\textcolor{rqup}{$\uparrow$} & 90.95 & 85.07 & 90.95 & \textbf{95.93} & 85.07 & 92.31 & 84.16 \\
\multirow{-5}{*}{\textcolor{rqgroup}{\textbf{Overall}}} & API\,\textcolor{rqdown}{$\downarrow$} & 86.7 & 134.8 & 160.2 & 146.1 & 160.2 & 177.1 & 151.1 \\
\bottomrule
\end{tabularx}
\end{table}

\subsection{Benchmark Construction}
\label{sec:construction}

\paragraph{Ecosystem and Change Mining}

From the top 200 GitHub organizations ranked by total repository stars (Gitstar Ranking, June 5, 2026), we select 103 active software ecosystems supporting a common product, platform, or technology. We collect PRs merged since January 1, 2024 in non-archived repositories and identify explicit links to other repositories within the same ecosystem. Among 1,729,171 PRs, 109,233 contain such links. We group linked changes across repositories into candidates. After deduplication and checks for PR availability, test-related changes, and Linux compatibility, 4,437 candidate groups remain.

\paragraph{Case Selection and Composition}
\label{sec:composition}

Figure~\ref{fig:benchmark-selection}(a) summarizes the selection process. We focus on 2,188 candidate groups whose latest PR was merged on or after June 1, 2025. Manual review retains 635 groups that address a single feature or bug fix and require substantive changes across repositories. Executable validation and final review yield 192 eligible cases, each with at least one \fpt{} test per target repository that fails before the reference changes and passes afterward. To balance bug fixes and features, we retain all 60 bug-fix cases and select 60 of the 132 feature cases, considering the diversity of required behaviors. The resulting 120 tasks cover 41 ecosystems (Figure~\ref{fig:ecosystem-map}(b)). Detailed selection criteria and dataset distributions appear in Appendix~\ref{app:construction}.

\paragraph{Prompt Construction}
\label{sec:prompts}

We construct each prompt from the original issues and PR descriptions, preserving their wording wherever possible and combining related requirements into one cross-repository task. We clarify necessary behavior and edge cases without introducing additional implementation instructions. Appendix~\ref{app:prompt-examples} illustrates this process with a comparison between the original text and the final prompt.

\paragraph{Hidden-Test Construction and Review}
\label{sec:audit}

We construct hidden tests by extracting test changes from the reference patches. During review, we find that some tests impose implementation-specific constraints beyond the task requirements, rejecting otherwise correct solutions. DeepSWE reports similar findings~\citep{huang2026deepswe}. We address these mismatches using two review rules, guided by the original issues, PR descriptions, and task prompts:
\begin{enumerate}[leftmargin=*,nosep]
    \item \textbf{Relax implementation-specific constraints.} Some tests require specific private helper names, files at fixed paths, or exact text in error messages, even though the task does not require these choices. We remove these restrictions while preserving required behavior and existing interface contracts.
    \item \textbf{Remove unrequested functionality checks.} Some reference patches introduce additional functionality required by neither the original issue/PR nor the task prompt. We remove tests for these additions while retaining checks for required functionality and existing behavior.
\end{enumerate}
For example, a TensorDict test checks that an invalid argument raises \texttt{TypeError} and that the error message contains ``generator must be.'' We retain the exception-type check but remove the message-text restriction. Appendix~\ref{app:test-audit-example} illustrates both review rules. We verify that the revised test suites still reject the original code and accept the reference solutions.

\begin{figure}[!t]
\centering
\includegraphics[width=\linewidth]{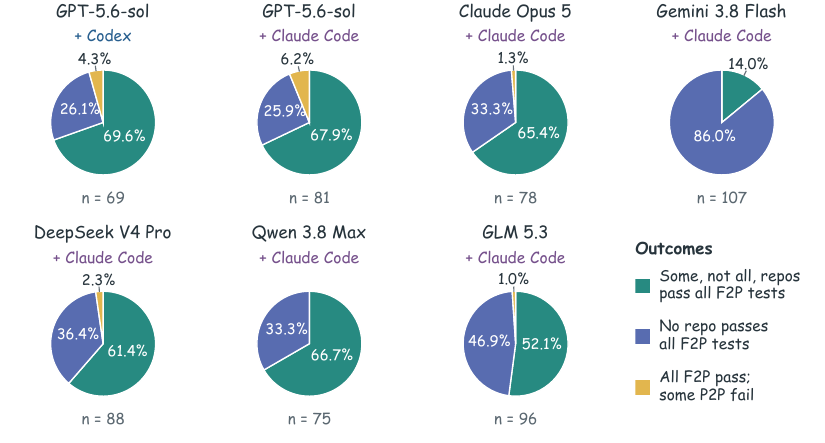}
\caption{Repository completion within unresolved tasks; $n$ is the number of unresolved tasks. Colors denote mutually exclusive outcomes.}
\label{fig:failure-decomposition}
\end{figure}

\subsection{Evaluation}
\label{sec:evaluation}

We evaluate the patched workspace $\mathcal{W}'_c$ using hidden tests supplied independently of agent edits (Appendix~\ref{app:test-isolation}). For each target repository $r \in \mathcal{T}_c$, let $F_{c,r}$ denote the fail-to-pass (\fpt{}) tests that fail on the original code and pass with the reference changes, and let $P_{c,r}$ denote the pass-to-pass (\ppt{}) regression tests that pass in both states.

Define $y_{c,r}(\mathcal{W}'_c)$ as 1 if every test in $F_{c,r} \cup P_{c,r}$ passes with no missing or invalid results, and 0 otherwise. A task is solved only when every target repository succeeds:
\begin{equation}
    \operatorname{Success}(c) = \prod_{r \in \mathcal{T}_c} y_{c,r}(\mathcal{W}'_c).
    \label{eq:task-success}
\end{equation}
The task success rate averages $\operatorname{Success}(c)$ over evaluated tasks. Repository-level and test-level results provide additional diagnostics.

\section{Experimental Setup}
\label{sec:setup}

We evaluate seven agent configurations, each pairing a coding-agent scaffold, an LLM, and a reasoning-effort setting. The scaffolds are Codex CLI (v0.147.0) and Claude Code (v2.1.139). Codex CLI is paired with GPT-5.6-sol (high), while Claude Code is paired with GPT-5.6-sol (high), Claude Opus 5 (high), Gemini 3.8 Flash (xhigh), DeepSeek V4 Pro (high), Qwen 3.8 Max~\citep{qwen38} (xhigh), and GLM 5.3 (xhigh). For each task, the agent receives one prompt and a historical ecosystem workspace, where it can inspect and modify all repositories. All configurations use the same harness-level permissions, timeout policy, and per-run resource limits.

We report results separately for bug fixes, features, and all tasks combined. \emph{Case-level success} is the primary metric and requires all target repositories to pass all \fpt{} and \ppt{} tests. \emph{Repository-level success} applies the same criterion to each target repository. \emph{F2P-complete} and \emph{P2P-preserved} report the percentage of repositories passing all corresponding tests, among those with such tests. Test-level pass rates appear in Appendix~\ref{app:test-level-rates}. API measures mean requests per task with recorded usage.

We ask:
\begin{description}[leftmargin=1.05cm,style=nextline,nosep]
    \item[RQ1] How effectively can coding agents complete cross-repository tasks?
    \item[RQ2] How does success vary with task type, repository count, and language diversity?
    \item[RQ3] How does joint execution compare with independent execution in each target repository under identical prompts?
\end{description}
For RQ3, we compare joint and independent execution using the Codex CLI--GPT-5.6-sol configuration on 89 cases (29 bug fixes and 60 features) whose prompts apply unchanged in both settings (Appendix~\ref{app:joint-independent}). The other 31 cases require repository-specific prompt adaptations and are excluded to keep the prompts identical. Selected trajectories in Appendix~\ref{app:trajectory-evidence} complement the quantitative results with comparisons across execution settings, agent scaffolds, and models.

\section{Results}
\label{sec:results}

\subsection{RQ1: Performance on Cross-Repository Tasks}

Table~\ref{tab:main-results} summarizes results on 120 tasks spanning 253 target repositories, with 2,815 \fpt{} and 22,139 \ppt{} tests. The best-performing configuration, \system{}, fully solves only 42.50\% of tasks; the other configurations achieve 10.83\%--37.50\%. Yet \system{} solves at least one target repository in 83.33\% of tasks, revealing a gap between repository-level progress and complete task resolution. We examine the unresolved tasks to understand where work remains incomplete and how these gaps arise.

\FloatBarrier

\paragraph{Functional completion within unresolved tasks.}
Figure~\ref{fig:failure-decomposition} examines the unresolved tasks. For every configuration except Gemini, most of these tasks (52.08\%--69.57\%) have at least one repository passing all F2P tests, but not all repositories do so. For Gemini, 85.98\% have no repository passing all F2P tests. Failures due solely to P2P tests are uncommon. Together with the high P2P preservation rates in Table~\ref{tab:main-results}, these results indicate that agents generally preserve existing tested behavior but often leave the requested functionality incomplete.

These results do not explain why agents leave work incomplete. We classify unresolved runs using a two-stage procedure based on final repository diffs and trajectory evidence; Appendix~\ref{app:failure-annotation} details the annotation rules and inter-annotator agreement. We group failures into the following three categories.

\textbf{Trajectory analysis locates where completion breaks down.}
Table~\ref{tab:failure-stages-configurations} groups unresolved runs into three categories. \emph{(1) Incomplete scope identification:} the agent leaves a necessary repository change unrecognized, or incorrectly decides that the repository needs no changes even after inspecting it. \emph{(2) Recognized work without delivery:} the agent identifies a necessary change or diagnoses the defect requiring it, but does not deliver the change. \emph{(3) Post-edit failure:} all target repositories receive substantive code changes, but the run still fails the required behavior or regression checks.

\FloatBarrier
\begin{table}[!t]
\centering
\begin{minipage}[t]{0.49\linewidth}
\vspace{0pt}
\footnotesize
\caption{Failure categories (\%).}
\label{tab:failure-stages-configurations}
\setlength{\tabcolsep}{2.5pt}
\renewcommand{\arraystretch}{1.12}
\begin{tabularx}{\linewidth}{@{}Xrrr@{}}
\toprule
\rowcolor{rqheader}
Configuration & Scope & Delivery & Post-edit \\
\midrule
GPT \textcolor{rqcodex}{+ Codex} & 37.68 & 2.90 & 59.42 \\
GPT \textcolor{rqcc}{+ CC} & 30.86 & 1.23 & 67.90 \\
Opus \textcolor{rqcc}{+ CC} & 25.64 & 6.41 & 67.95 \\
\rowcolor{rqdown!9}
\textcolor{rqdown}{\textbf{Gemini}} \textcolor{rqcc}{+ CC} & 37.38 & \textcolor{rqdown}{\textbf{52.34}} & 10.28 \\
DeepSeek \textcolor{rqcc}{+ CC} & 30.68 & 7.95 & 61.36 \\
Qwen \textcolor{rqcc}{+ CC} & 21.33 & 5.33 & 73.33 \\
GLM 5.3 \textcolor{rqcc}{+ CC} & 30.21 & 9.38 & 60.42 \\
\bottomrule
\end{tabularx}
\end{minipage}\hfill
\begin{minipage}[t]{0.48\linewidth}
\vspace{0pt}
\footnotesize
\caption{Joint versus Independent execution.}
\label{tab:joint-independent}
\setlength{\tabcolsep}{2pt}
\renewcommand{\arraystretch}{1.08}
\begin{tabularx}{\linewidth}{@{}lXrr@{}}
\toprule
\rowcolor{rqheader}
Group & Metric & \textcolor{rqcodex}{Joint} & \textcolor{rqdown}{Independent} \\
\midrule
 & Bugfix & 34.48 & \textbf{44.83} \\
 & Feature & \textbf{43.33} & 31.67 \\
\multirow{-3}{*}{\textcolor{rqgroup}{\shortstack[l]{\textbf{Task}\\\textbf{success (\%)}}}} & \cellcolor{rqcase}Overall & \cellcolor{rqcase}\textbf{40.45} & \cellcolor{rqcase}35.96 \\
\specialrule{0.3pt}{0.6pt}{0.6pt}
 & Repo & \textbf{61.78} & 60.73 \\
 & \fpt{} & \textbf{64.40} & \textbf{64.40} \\
\multirow{-3}{*}{\textcolor{rqgroup}{\shortstack[l]{\textbf{Repository}\\\textbf{outcomes (\%)}}}} & \ppt{} & \textbf{89.63} & 85.37 \\
\specialrule{0.3pt}{0.6pt}{0.6pt}
 & API & \textbf{94.7} & 296.6 \\
\bottomrule
\end{tabularx}
\end{minipage}
\end{table}

For the Codex CLI--GPT-5.6-sol configuration, the shares in Table~\ref{tab:failure-stages-configurations} are 37.68\%, 2.90\%, and 59.42\%, respectively. Post-edit failure is most common. Some runs fully fix one repository but make only part of the necessary changes in another, leaving the problem unresolved. Others connect a caller to new functionality in another repository but change the caller's existing behavior, causing regression tests to fail. A subtler failure occurs in Sentry/Symbolicator: the agent makes the sender and receiver use an exception-name string, and its tests also use strings, although the task requires structured exception information. Both implementations and the added tests share the same incorrect assumption, allowing local tests to pass without delivering the required behavior (Appendix~\ref{app:codex-joint-cases}).

Some Delivery failure cases show that even with a clear task scope, agents may still fail to complete the implementation after many API calls. For example, DeepSeek makes 769 API calls on the Prettier/yaml-unist-parser task, explicitly planning changes to both repositories early on but continuing to work on only one, ultimately leaving Prettier unchanged. Gemini makes 562 API calls on the Graphon/Dify task, completing Graphon before moving on to investigate Dify but never delivering the Dify changes.

\textbf{Same scaffold, different models.}
With Claude Code fixed, Table~\ref{tab:failure-stages-configurations} shows distinct patterns among each model's unresolved runs. In 52.34\% of Gemini's failures, it identifies the necessary work but leaves changes undelivered; another 37.38\% involve incomplete scope identification. Within the former category, 82.14\% remain in investigation, solution planning, or preparatory checks rather than implementing the requested behavior; 17.86\% begin implementing in some repositories but leave recognized work in others undone. In these failed runs, Gemini often remains in analysis rather than proactively implementing the required changes. In Elastic Agent/Fleet Server, for example, Gemini explicitly plans client-side request compression and server-side support, but continues investigating configuration and request handling without implementing either change (Appendix~\ref{app:model-comparison}, Figure~\ref{fig:fleet-case-study}).

Recognized work without delivery is less common for DeepSeek (7.95\%), Qwen (5.33\%), and GLM (9.38\%): their omissions more often involve failing to recognize the full modification scope. Post-edit failures dominate for Opus (67.95\%) and Qwen (73.33\%), as well as GLM. Unlike Gemini, which frequently leaves recognized work unfinished, these models often locate and modify the relevant repositories, yet still struggle to make those changes jointly satisfy the request. Appendix~\ref{app:model-comparison} presents selected comparative trajectories.

\begin{figure}[!t]
\centering
\includegraphics[width=\linewidth]{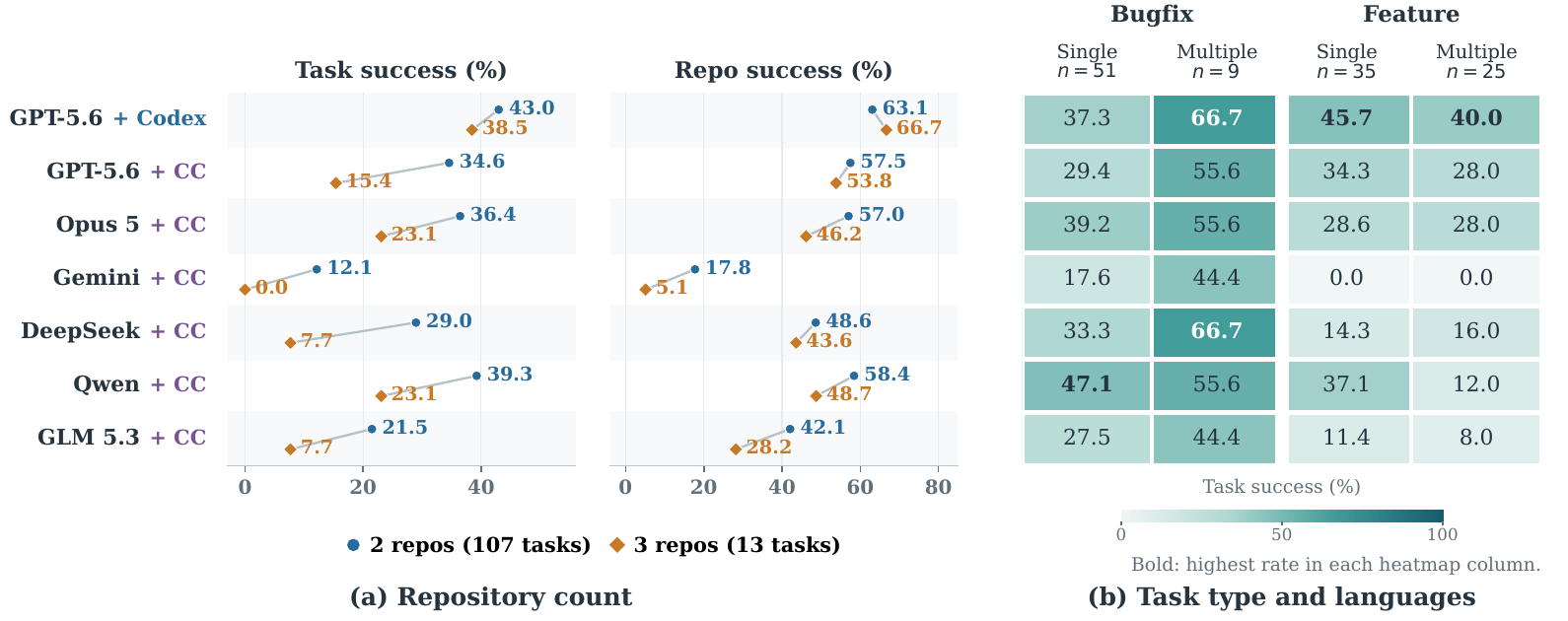}
\caption{(a) Task and repository success by target-repository count. Connected points compare separate task groups. (b) Task success by intent and language diversity; $n$ denotes the number of tasks. Language families describe reference changes. All rates are percentages.}
\label{fig:task-attributes}
\end{figure}

\textbf{Same model, different scaffolds.}
Table~\ref{tab:main-results} shows that, with GPT-5.6-sol fixed, moving from the Codex CLI scaffold to Claude Code reduces task success from 42.50\% to 32.50\%, while mean recorded API requests per task rise from 86.7 to 134.8. The Codex CLI pairing is more effective in this evaluation; the higher request count does not correspond to higher task completion. Appendix~\ref{app:scaffold-comparison} provides a case comparison of the same task under the two scaffolds.

\subsection{RQ2: Task Intent, Repository Scope, and Language Diversity}
\label{sec:rq2-results}

We compare task type, language diversity, and repository count using Table~\ref{tab:main-results} and Figure~\ref{fig:task-attributes}, then examine failure behaviors. Language families describe reference changes (Appendix~\ref{app:provenance-labels}).

\FloatBarrier

\paragraph{Model performance varies by task type and language mix.}
The Claude Code--Qwen configuration outperforms the Codex CLI--GPT-5.6-sol configuration on bug fixes but falls behind on features (Table~\ref{tab:main-results}). Figure~\ref{fig:task-attributes}(b) further shows that all seven configurations achieve higher success on multiple-family bug fixes than on single-family bug fixes, but feature tasks do not follow the same pattern. For example, Qwen leads the Claude Code configurations on single-family features at 37.14\%, but falls to 12.00\% on multiple-family features, below GPT and Opus at 28.00\%. Opus remains comparatively stable across the two feature groups. Language count alone is therefore insufficient to judge task difficulty; task type and the specific changes required also matter.

\paragraph{Performance by Repository Count.}
In our experiments, all configurations have lower task success on three-repository tasks (Figure~\ref{fig:task-attributes}(a)). This gap does not necessarily arise from overlooking more repositories. For the Codex CLI--GPT-5.6-sol configuration, repository success rises from 63.08\% on two-repository tasks to 66.67\% on three-repository tasks, while task success falls from 42.99\% to 38.46\%. Among its failed three-repository tasks, 87.50\% modify all targets and 62.50\% solve two of them. These results suggest that this configuration usually covers the repositories requiring changes, but coordinating the required adaptations may become more difficult as the number of repositories increases. A three-repository Sentry task illustrates this distinction in a Claude Code--DeepSeek run: the agent modifies all three targets and completes the PHP SDK and Laravel changes, but adds an integer-only configuration node in Symfony, rejecting the \texttt{null} setting explicitly allowed by the request (Appendix~\ref{app:codex-joint-cases}, Figure~\ref{fig:sentry-configuration-case-study}).

\subsection{RQ3: Joint versus Independent Execution}
\label{sec:rq3-results}

\begin{figure}[t]
\centering
\includegraphics[width=\linewidth]{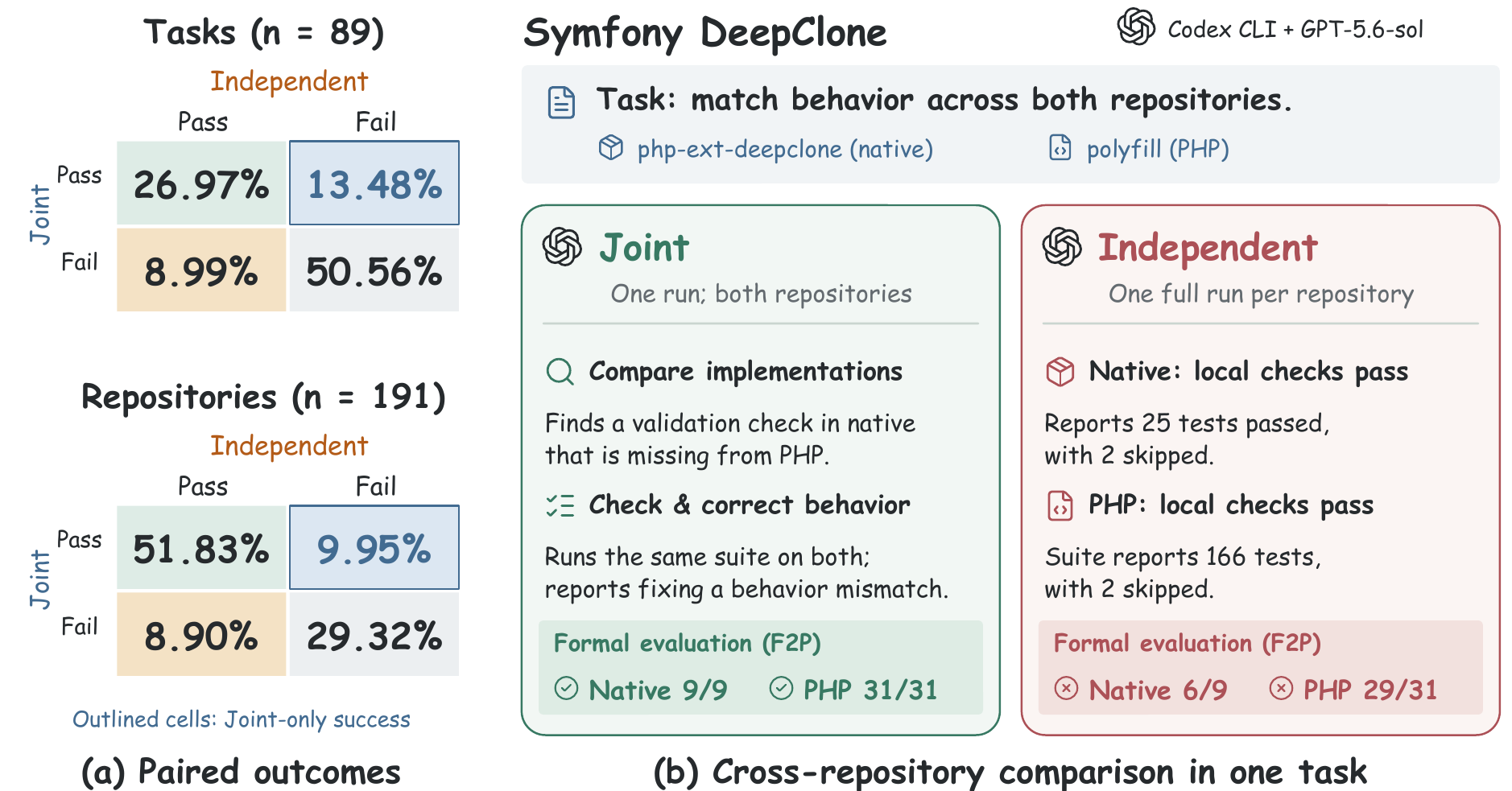}
\caption{(a) Paired task and repository outcomes (\%); off-diagonal cells show success reversals. (b) A Symfony task where Joint compares related implementations and both repositories pass, while Independent leaves failures in both. F2P scores show passed/total hidden checks, not local tests.}
\label{fig:scope-transitions}
\end{figure}

\paragraph{Independent execution does not consistently improve task success.}
Table~\ref{tab:joint-independent} compares joint and independent execution of the Codex CLI--GPT-5.6-sol configuration under identical prompts on 89 tasks. Joint execution solves 40.45\%, versus 35.96\% when every independent repository run must succeed. Yet 22.47\% of tasks change outcome in either direction (Figure~\ref{fig:scope-transitions}). Independent execution improves bug-fix success from 34.48\% to 44.83\%, but reduces feature success from 43.33\% to 31.67\%. This contrast suggests that narrowing the work scope affects bug fixes and features differently. For bug fixes, a narrower scope can help agents focus on locating and repairing defects in the current repository. Features often require repositories to jointly implement new behavior; separating them may remove context needed to understand interface contracts and compare related implementations, making coordinated implementation harder.

Independent execution uses 296.6 API requests per task on average, compared with 94.7 for Joint, because each target receives a separate full run. Despite using 3.13 times as many requests, independent execution does not improve overall task success.

\paragraph{Independent execution reduces each run's scope and workload.}
Among repositories that fail in Joint execution, 60.00\% of those left unmodified succeed independently, compared with only 9.43\% of those already modified (Table~\ref{tab:scope-recovery} in Appendix~\ref{app:joint-independent}). This suggests that a separate run offers much less improvement when an implementation has already been attempted but remains incorrect. In the Sentry Go/Python/Ruby task, the Joint run completes only the Go SDK, whereas separate runs complete all three SDKs under the same prompt (Appendix~\ref{app:scope-case-comparison}, Figure~\ref{fig:sentry-case-study}). Additionally, in the MUI Strict Mode bug-fix task, the Joint run recognizes that both repositories require changes but ends after repairing Base UI and resolving its testing setup, forgetting to modify Material UI. Separate runs repair both repositories. This case suggests that handling a larger workload within a joint run may lead the agent to forget required changes in another repository.

\paragraph{Cross-repository context supports both implementation and verification.}
Figure~\ref{fig:scope-transitions} shows that 13.48\% of tasks and 9.95\% of repositories succeed only in Joint execution. Trajectories illustrate two uses of cross-repository context. First, it supplies facts needed for implementation. In Ansible, the Joint run uses one repository's implementation to guide compatible changes in the other. The independent run instead makes incorrect assumptions about the other repository, producing incompatible code (Appendix~\ref{app:scope-case-comparison}, Figure~\ref{fig:ansible-case-study}). Second, it provides behavioral references for verification. In Symfony, the two repositories must implement the same behavior. The Joint run compares their implementations and corrects a difference, allowing both repositories to pass (Figure~\ref{fig:scope-transitions}(b)); the independent runs leave errors unresolved in both (Appendix~\ref{app:scope-case-comparison}, Figure~\ref{fig:symfony-case-study}).

\section{Related Work}
\label{sec:related}

\paragraph{Coordination in software ecosystems.}
Dependencies between software projects extend development work beyond a single project~\citep{blincoe2015ecosystems,decan2019empirical}. Reference Coupling identifies technical dependencies through cross-project references in development records~\citep{blincoe2019reference}. \citet{ma2017developers} study how developers trace bug causes across projects and coordinate upstream and downstream repairs. An industrial study further documents how teams maintaining interdependent components coordinate interface changes, work progress, and priorities~\citep{begel2008effecting}. These studies examine project dependencies and human coordination; \benchmark{} turns real cross-repository changes into executable tasks, evaluating whether coding agents can identify the required change scope and fulfill a shared request across repositories.

\paragraph{Software-engineering agents and benchmarks.}
RepoBench and CrossCodeEval~\citep{liu2024repobench,ding2023crosscodeeval} evaluate code completion~\citep{hindle2016naturalness,raychev2014code} using cross-file repository context~\citep{zhang2023repocoder,wu2024repoformer,le2025impacts,wang2025grace,zhao2025completion,wang2026greprag}. SWE-bench evaluates real GitHub issue resolution through executable tests~\citep{jimenez2024swe}, a setting studied by SWE-agent, Agentless, AutoCodeRover, and OpenHands~\citep{yang2024swe,xia2025demystifying,zhang2024autocoderover,wang2025openhands}. Subsequent benchmarks broaden evaluation along several dimensions~\citep{yang2024swebench,zhang2026swe,badertdinov2026swe,badertdinov2026swev2}. Multi-SWE-bench, SWE-PolyBench, and SWE-bench Multilingual extend language coverage beyond Python~\citep{zan2026multi,rashid2025swe,yang2026swe}. SWE-Bench Pro and SWE-EVO emphasize long-horizon engineering and software evolution~\citep{deng2025swe,le2025swe}. DeepSWE introduces original long-horizon tasks with functional verifiers, while ProgramBench requires reconstructing programs from executables and documentation~\citep{huang2026deepswe,yang2026programbench}. Broader language coverage and longer tasks expand agents' work within a project; they do not themselves evaluate whether agents can complete coordinated changes across repositories.

\paragraph{Cross-repository evaluation.}
BeyondSWE's CrossRepo tasks require agents to resolve an issue in a target repository using code and solutions from external repositories~\citep{chen2026beyondswe}. \benchmark{} focuses on implementing one shared feature or fix across multiple scored target repositories. Related repositories can provide useful context, but they also contain required changes that must be delivered and verified. Task success therefore requires all target repositories to pass their required tests.

\section{Conclusion}

We introduce \benchmark{}, a benchmark of 120 real-world tasks requiring coordinated changes across repositories to implement one feature or fix. Across the seven evaluated configurations, the highest task success rate is 42.50\%. Agents miss necessary changes, leave recognized work unfinished, or modify the required repositories without fully satisfying the request. Independent execution can recover omitted work but does not improve overall task success in our paired comparison. Joint execution can provide information from related repositories that helps agents implement and verify changes. \benchmark{} shifts the focus from progress in individual repositories to whether changes across repositories jointly fulfill the task requirements.

\subsection*{AI Use Statement}
We used generative AI tools to assist with manuscript writing and editing, code and test development, and result analysis. The authors reviewed the AI-assisted work and are responsible for the content.

\subsection*{Ethics Statement}
The benchmark is constructed from public open-source repositories and public development records. The released artifact preserves licenses and attribution, excludes credentials and private data, and documents responsible use. Agent-generated patches may contain vulnerabilities or regressions and should not be deployed without human review.

\subsection*{Reproducibility Statement}
The artifact includes instance metadata, repository and base-commit identifiers, provenance records, prompts, hidden-test patches, environment definitions, scoring manifests, agent patches where licensing permits, and scripts for reproducing aggregate metrics. Section~\ref{sec:benchmark} describes construction and evaluation, and Section~\ref{sec:setup} defines the experimental conditions and metrics.

\bibliography{iclr2027_conference}
\bibliographystyle{iclr2027_conference}

\appendix
\section{Benchmark Construction Details}
\label{app:construction}

\subsection{Candidate Discovery and Time Scope}

The ecosystem screening in Section~\ref{sec:benchmark} checks each organization's highest-ranked repositories. We exclude organizations with fewer than two non-archived repositories in this group or no activity within the preceding 24 months.

Cross-repository PR and issue links identify candidate change groups. Test-related paths, such as \texttt{tests/}, \texttt{spec/}, and \texttt{\_test}, provide an initial screening signal; behavioral coverage is checked during executable validation. Candidates must be evaluable on Linux. Record-level deduplication uses the ecosystem, source repository, source PR, and involved-repository set.

The review cutoff applies to the latest merge timestamp among the involved PRs, so individual constituent PRs may predate June 1, 2025. The 109,233 cross-referencing PR records count linked activity rather than distinct tasks; link direction does not imply architectural dependency direction.

\subsection{Semantic Review, Case Assembly, and Selection}

Reviewers examine PR descriptions, linked discussions, changed files, and tests to establish whether the records address one feature or fix and require substantive changes in multiple repositories. An incidental citation does not suffice. Of the 635 candidates retained for environment construction, 240 are runnable and 192 satisfy the per-target \fpt{} requirement.

Record deduplication is distinct from assembling a shared change chain. When multiple candidate records describe the same request, their relevant requirements and repository changes are consolidated rather than released as separate copies. For example, the gRPC timeout interoperability case combines two candidates sharing the same Java PR with two related Go PRs. The construction record regenerates the combined Go reference and test patches and reruns Base/Gold validation. This merge is justified by a common interoperability requirement, not merely by similar titles or overlapping repositories.

The eligible pool contains 60 bug fixes and 132 features. To balance task types while preserving a diverse range of ecosystems and cross-repository changes, we retain all bug fixes and select 60 features, including all 13 three-repository cases and 47 two-repository cases. Feature selection favors substantive changes involving runtime behavior, protocols, schemas, or SDK integration over narrow field forwarding or mechanical updates. Within heavily represented ecosystems, we reduce repeated coverage of similar change types while preserving ecosystem coverage, yielding a balanced release of 120 tasks.

\subsection{Dataset Distributions}
\label{app:dataset-statistics}

\paragraph{Coordination patterns.} All 120 tasks require coordinated changes across repositories to fulfill a shared feature or bug-fix request. Table~\ref{tab:coordination-patterns} summarizes two task-level semantic patterns. \emph{Producer--consumer dependency} accounts for 96 tasks (80.0\%): one repository provides capabilities, interfaces, schemas, service behavior, or build artifacts that other repositories consume or adapt to; this category also includes compatibility between protocol endpoints, as illustrated in Figure~\ref{fig:grpc-case-study}. \emph{Parallel propagation} accounts for 24 tasks (20.0\%): multiple repositories implement the same behavior or fix, as in language SDKs, providers, or native and polyfill implementations. Across both patterns, agents must identify the required repository scope and carry the shared request through all necessary implementations. Related implementations can also provide behavioral references and guide adaptations in other repositories, as illustrated by the Sentry SDK case in Appendix~\ref{app:codex-joint-cases}.

\begin{table}[!ht]
\centering
\small
\caption{Cross-repository coordination patterns across 120 tasks.}
\label{tab:coordination-patterns}
\setlength{\tabcolsep}{5pt}
\renewcommand{\arraystretch}{1.12}
\begin{tabularx}{\linewidth}{@{}Xrr@{}}
\toprule
\rowcolor{rqheader}
Coordination pattern & Tasks & Share (\%) \\
\midrule
Producer--consumer dependency & 96 & 80.0 \\
Parallel propagation & 24 & 20.0 \\
\midrule
Total & 120 & 100.0 \\
\bottomrule
\end{tabularx}
\end{table}

\begin{figure}[htbp]
    \centering
    \includegraphics[width=\linewidth]{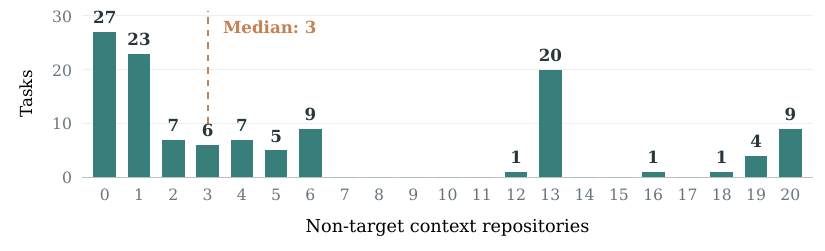}
    \caption{Non-target context repositories available per task across 120 historical workspaces. Bar labels show task counts; the dashed line marks the median. Scored target repositories are excluded.}
    \label{fig:dataset-statistics}
\end{figure}

\paragraph{Workspace scope.} Context repositories are available for reference but are not scored targets. In Figure~\ref{fig:dataset-statistics}, the horizontal axis gives their number in a task's workspace, and bar heights count tasks with that number. For example, 27 tasks have no additional context repositories, while 93 have at least one. The range is zero to 20, with a median of three per task.

\begin{figure}[htbp]
    \centering
    \includegraphics[width=\linewidth]{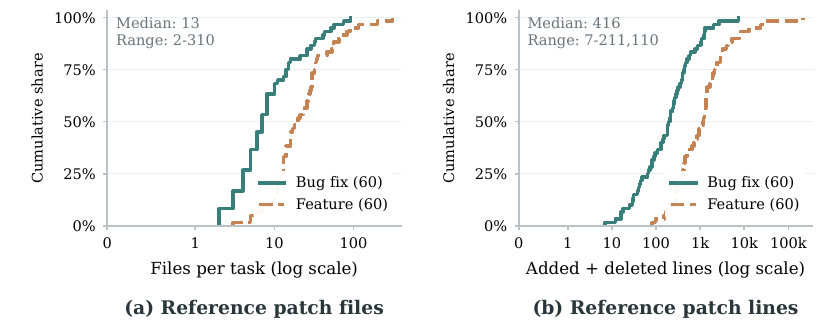}
    \caption{Cumulative reference-patch sizes by task type. Horizontal axes are logarithmic above one and linear near zero.}
    \label{fig:dataset-patches}
\end{figure}

\paragraph{Reference-patch size.} Figure~\ref{fig:dataset-patches} compares the number of files and lines changed by reference patches for bug fixes and features. At each horizontal-axis value, the curve gives the proportion of tasks with no more than that many changed files or lines; a curve farther to the right therefore indicates larger patches. Feature patches are generally larger than bug-fix patches. Across all tasks, the medians are 13 files and 416 lines. Counts include production code, tests, generated files, documentation, and configuration. Line counts sum additions and deletions; binary files contribute only to file counts. Patch size describes the changes, not task difficulty.

\begin{figure}[htbp]
    \centering
    \includegraphics[width=\linewidth]{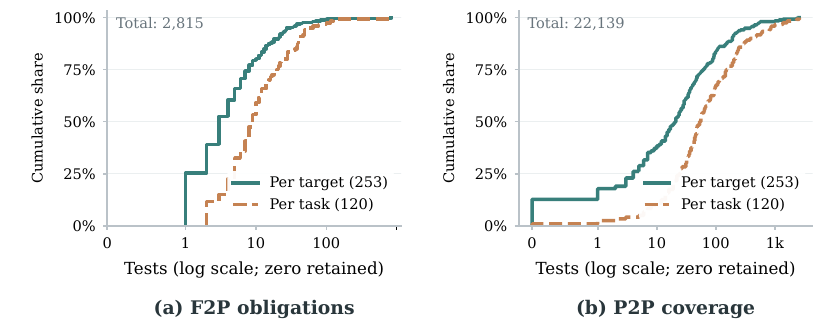}
    \caption{Cumulative test counts per target repository and per task. Horizontal axes retain zero counts and use the same scale as Figure~\ref{fig:dataset-patches}.}
    \label{fig:dataset-tests}
\end{figure}

\paragraph{Test coverage.} Figure~\ref{fig:dataset-tests} summarizes 2,815 \fpt{} and 22,139 \ppt{} tests across 253 target-repository instances, using the scoring manifests. The solid curves count tests within each target repository; the dashed curves sum tests across all targets in a task. As above, the vertical axis gives the proportion with no more than the test count on the horizontal axis, not an agent success rate. The per-task curves lie farther to the right because each task combines tests from multiple repositories.

\subsection{Prompt Provenance and Language Labels}
\label{app:provenance-labels}

Prompt construction prioritizes sentences from the original issues and PRs. It removes templates, repeated passages, links, and irrelevant material while retaining observable requirements, reproduction conditions, public interfaces, compatibility constraints, and necessary edge cases. Case records map each prompt paragraph to its source and identify any consolidation or necessary clarification. The Gold implementation is not expanded into a step-by-step solution; private symbols, file layouts, and algorithms are not added merely because that implementation uses them. Test-required behavior must be supported by the public demand and expressed sufficiently for a correct implementation, rather than hidden behind an undisclosed naming or representation choice.

Language-family labels use reference production-code changes, excluding tests, documentation, configuration, conventional build-support files, and generated bundles. A task spanning multiple families may do so within a repository or across repositories. The resulting 86/34 split describes language diversity in the required work, not a controlled intervention on language.

\subsection{Prompt Construction Example}
\label{app:prompt-examples}

One author constructed the prompts, and two other authors independently reviewed each prompt against its source issues and pull requests to verify that all required behavior was preserved and no implementation-specific instructions were introduced. Disagreements were resolved through discussion.

\begin{figure}[p]
    \centering
    \includegraphics[width=\linewidth]{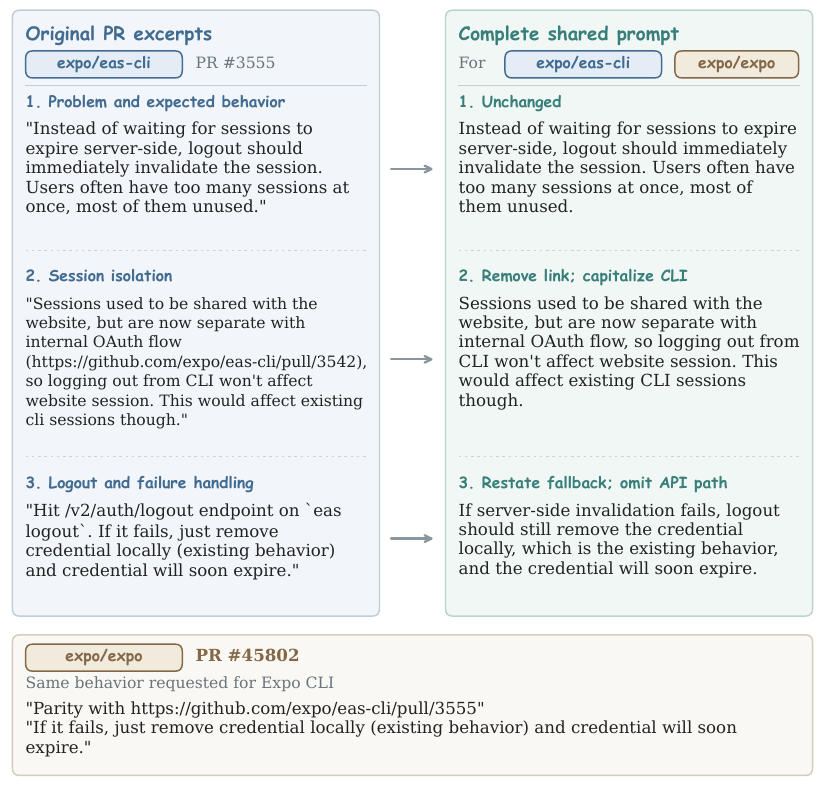}
    \caption{Prompt construction for the Expo logout task. Left: wording from eas-cli\#3555. Right: the complete shared prompt. Bottom: expo\#45802 establishes the same requirement in the second repository.}
    \label{fig:prompt-construction}
\end{figure}

Figure~\ref{fig:prompt-construction} compares the source passages with the complete prompt for the Expo logout task. \href{https://github.com/expo/eas-cli/pull/3555}{eas-cli\#3555} supplies the problem description and behavioral requirements. \href{https://github.com/expo/expo/pull/45802}{expo\#45802} explicitly requests parity with that change and repeats the failure-handling requirement, establishing that the request also applies to Expo CLI. The shared prompt consolidates these overlapping requirements rather than repeating the same behavior once per repository.

The first paragraph is unchanged from the PR. The second removes a PR link and capitalizes ``CLI.'' The third is reworded to preserve the shared logout and failure-handling requirements while omitting the specific command and endpoint, leaving agents to determine how to implement the behavior from each repository's code.

\subsection{Historical Workspace Snapshots}

Each target repository uses the base commit of its corresponding PR. Context repositories use snapshots no later than the earliest target base to prevent future-information leakage. Gold patches and hidden tests are not exposed to agents.

\subsection{Evaluator Test Isolation and Patch Conflicts}
\label{app:test-isolation}

Agent changes are evaluated on clean Base snapshots, with evaluator-owned tests supplied separately. When a hidden-test patch targets a public test file that the agent has also edited, the evaluator can materialize the hidden version from the clean Base in a temporary directory and copy only the patch's target files into the evaluation workspace. Test-patch assembly therefore does not depend on whether the agent's public-test edits happen to match the patch context. Files outside the evaluator patch targets are not changed by this overlay.

For profiles that relocate upstream tests into separate hidden files, the evaluator materializes the source test from Base and installs it at the configured hidden path. Where compilation would otherwise encounter duplicate test declarations, the enclosing test type is renamed while assertions and test bodies are preserved. Overlapping evaluator patches are applied in their configured order so that a later patch does not discard an earlier evaluator layer. Patch-injection failures are recorded and cannot produce a successful evaluation. These mechanisms address test-file and declaration conflicts, not behavioral acceptance criteria or an agent's production-code implementation.

Section~\ref{app:test-audit-example} documents revisions to test acceptance criteria separately from these test-installation mechanisms.

\subsection{Examples of Requirement-Aligned Test Review}
\label{app:test-audit-example}

\subsubsection{Relax Implementation-Specific Constraints}
\label{app:test-wording}

\begin{figure}[!ht]
    \centering
    \includegraphics[width=\linewidth]{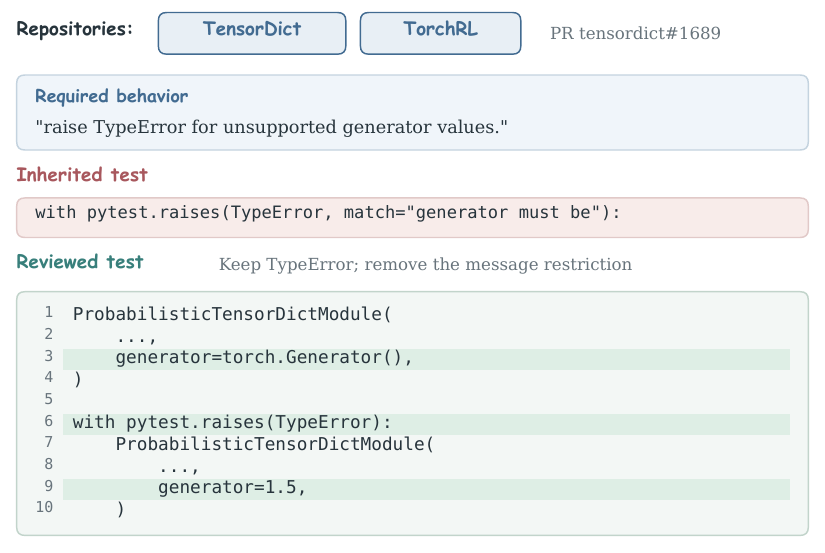}
    \caption{Preserve the required exception without fixing its wording. Excerpts normalize formatting and quotation marks; line numbers are local and ellipses mark omissions.}
    \label{fig:test-review-wording}
\end{figure}

\paragraph{Request and scope.}
The TensorDict/TorchRL task adds a local random generator for probabilistic sampling and forwards the option through the actor interface. The TensorDict PR explicitly lists \texttt{invalid type raises TypeError} among its requirements (\href{https://github.com/pytorch/tensordict/pull/1689}{tensordict\#1689}). The final prompt likewise states:
\begin{requirementquote}
Preserve the existing behavior when \texttt{generator=None}, and raise \texttt{TypeError} for unsupported generator values.
\end{requirementquote}
Neither statement specifies the error message.

\paragraph{Inherited test.}
The invalid-input test passes a floating-point value and checks both the exception type and the phrase \texttt{generator must be} (Figure~\ref{fig:test-review-wording}).

\paragraph{Audited test.}
The test first confirms that a supported generator is accepted, then checks that the same call with an unsupported value raises the required exception. Removing the message match permits different explanations of the same input error.

\subsubsection{Remove Unrequested Functionality Checks}
\label{app:test-unrequested}

\begin{figure}[!ht]
    \centering
    \includegraphics[width=\linewidth]{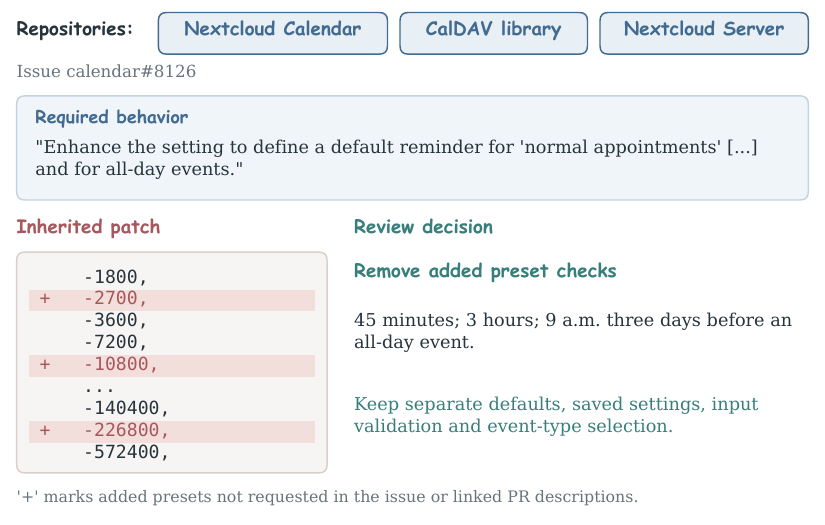}
    \caption{Remove unrequested preset expansion. Excerpts normalize formatting and quotation marks; ellipses mark omissions.}
    \label{fig:test-review-scope}
\end{figure}

\paragraph{Request and scope.}
The Nextcloud task spans Calendar, its CalDAV client library, and Server. Previously, one default reminder applied to both timed and all-day events. The issue requests separate settings:
\begin{requirementquote}
Enhance the setting to define a default reminder for "normal appointments" (not sure what would be a good name here) and for all-day events.
\end{requirementquote}
This passage is from \href{https://github.com/nextcloud/calendar/issues/8126}{calendar\#8126}, linked by \href{https://github.com/nextcloud/calendar/pull/8144}{calendar\#8144}, \href{https://github.com/nextcloud/cdav-library/pull/1012}{cdav-library\#1012}, and \href{https://github.com/nextcloud/server/pull/59517}{server\#59517}. The prompt specifies separate settings, their storage and propagation, input validation, and selection according to event type. It does not require additional built-in reminder presets.

\paragraph{Inherited test.}
Alongside the requested settings, the reference patch expands the preset lists. The inherited test patch adds \texttt{-2700}, \texttt{-10800}, and \texttt{-226800} to the expected arrays in \texttt{defaultAlarmProvider.test.js} (Figure~\ref{fig:test-review-scope}). The first two additions introduce 45-minute and three-hour presets. The third introduces a reminder at 9:00 a.m.\ three days before an all-day event. These additions are present in the reference implementation, but are not requested in the issue or linked PR descriptions.

\paragraph{Audited scope.}
We remove the preset-expansion hunks from the hidden patch rather than require these three additional choices. This does not remove support for user-configured offsets or the requirement to keep separate timed-event and all-day defaults. Tests for the DAV properties, client-model fields, saved settings, input validation, and event-type-specific behavior remain. The preset test file remains in the test profile; existing preset coverage is retained without making the added choices a task requirement.

\paragraph{Validation.}
Both revised cases retain failing target behavior on Base and pass with Gold. The TensorDict/TorchRL case has 21 \fpt{} and 580 \ppt{} checks; the Nextcloud case has 53 \fpt{} and 223 \ppt{} checks.

Each test revision was first proposed by one author and then independently assessed by two other authors with reference to the task requirements and the original issues/PRs. Revisions were accepted only when both reviewers judged that they removed implementation-specific or unrequested constraints without weakening the required behavior. Any disagreement was settled through discussion.

\paragraph{Revision coverage.}
Across the 120 tasks, requirement-aligned test revisions affected 88 tasks. Implementation-specific constraints were relaxed in 87 tasks, and unrequested functionality checks were removed in 16 tasks; 15 tasks involved both categories. These are task-level counts, not counts of individual tests or assertions, and exclude changes limited to test installation, fixtures, or execution environments.

\section{Supplementary Evaluation Results}

\subsection{Joint and Independent Repository Comparison}
\label{app:joint-independent}

For the 89 prompt-equivalent cases in Section~\ref{sec:setup}, \textbf{Joint} exposes the complete ecosystem workspace to one run, while \textbf{Independent} exposes only one target repository to each repository-specific run. Both use the same agent configuration, tool permissions, and per-run limits, but Independent receives one full run per target repository. This follows the conventional single-repository benchmark setup, in which each repository-level task receives its own run with the full per-run budget. The comparison therefore matches per-run limits, not the total budget available for each multi-repository task. Our aim is to examine whether solving repositories separately alleviates cross-repository difficulties, not to isolate the effect of execution scope under equal total budgets.

We pair outcomes for the same repository within the same case. Repository success requires all \fpt{} and \ppt{} tests to pass. At case level, Joint succeeds when one run resolves every target repository; Independent-all succeeds when every repository-specific run succeeds. Table~\ref{tab:joint-independent-pairs} reports the paired outcome percentages underlying Section~\ref{sec:rq3-results}.

\begin{table}[!ht]
\centering
\small
\caption{Paired Joint--Independent outcomes.}
\label{tab:joint-independent-pairs}
\setlength{\tabcolsep}{5pt}
\renewcommand{\arraystretch}{1.12}
\begin{tabularx}{\linewidth}{@{}Xrr@{}}
\toprule
\rowcolor{rqheader}
Outcome & Cases (\%) & Repositories (\%) \\
\midrule
Both succeed & 26.97 & 51.83 \\
\textcolor{rqcodex}{Joint} only & 13.48 & 9.95 \\
\textcolor{rqdown}{Independent} only & 8.99 & 8.90 \\
Neither succeeds & 50.56 & 29.32 \\
\bottomrule
\end{tabularx}
\end{table}

\paragraph{Recovery among Joint-failed repositories.}
Table~\ref{tab:scope-recovery} groups the 73 repositories that fail in Joint by whether the final Joint diff records a modification. This grouping does not use the Independent outcome, and a recorded modification does not imply a correct implementation. Recovery requires all \fpt{} and \ppt{} tests to pass in Independent. Because the groups differ in task composition, their recovery-rate difference is descriptive, not a causal effect of assigning a separate run.

\begin{table}[!ht]
\centering
\small
\caption{Independent recovery among Joint-failed repositories.}
\label{tab:scope-recovery}
\setlength{\tabcolsep}{5pt}
\renewcommand{\arraystretch}{1.12}
\begin{tabularx}{\linewidth}{@{}Xrrr@{}}
\toprule
\rowcolor{rqheader}
\textcolor{rqcodex}{Joint} modification & $n$ & \textcolor{rqdown}{Independent} pass & Recovery (\%) \\
\midrule
None & 20 & 12 & 60.00 \\
Present & 53 & 5 & 9.43 \\
\bottomrule
\end{tabularx}

\end{table}

\paragraph{Reversals by task intent.}
Among the 29 bug fixes, nine cases succeed in both settings, one only in Joint, four only in Independent-all, and 15 in neither. Among the 60 features, the corresponding counts are 15, 11, four, and 30. The opposite net directions should not be generalized beyond this sample.

\subsection{Test-Level Pass Rates}
\label{app:test-level-rates}

Tables~\ref{tab:test-level-rq1} and~\ref{tab:test-level-rq3} report the proportion of individual tests passed, computed as total passed tests divided by total tests in each group. Unlike the repository-level metrics in the main text, these rates capture partial progress within repositories and weight repositories by their test counts.

Configuration abbreviations in Table~\ref{tab:test-level-rq1} follow Table~\ref{tab:main-results}. Table~\ref{tab:test-level-rq3} uses the same 2,532 F2P and 18,051 P2P tests in each condition for the 89 paired RQ3 tasks.

\begin{table}[!ht]
\centering
\small
\caption{RQ1 test-level pass rates (\%).}
\label{tab:test-level-rq1}
\setlength{\tabcolsep}{4pt}
\renewcommand{\arraystretch}{1.12}
\begin{tabularx}{\linewidth}{@{}l*{6}{>{\raggedleft\arraybackslash}X}@{}}
\toprule
\rowcolor{rqheader}
& \multicolumn{2}{c}{Bugfix} & \multicolumn{2}{c}{Feature} & \multicolumn{2}{c}{Overall} \\
\cmidrule(lr){2-3}\cmidrule(lr){4-5}\cmidrule(lr){6-7}
Configuration & F2P & P2P & F2P & P2P & F2P & P2P \\
\midrule
GPT \textcolor{rqcodex}{+ Codex} & 75.00 & 99.89 & 83.57 & 98.01 & 81.74 & 98.84 \\
GPT \textcolor{rqcc}{+ CC} & 62.83 & 99.69 & 80.05 & 87.44 & 76.38 & 92.85 \\
Opus \textcolor{rqcc}{+ CC} & 74.33 & 98.80 & 39.82 & 87.96 & 47.18 & 92.75 \\
Gemini \textcolor{rqcc}{+ CC} & 31.00 & 99.77 & 2.62 & 95.13 & 8.67 & 97.18 \\
DeepSeek \textcolor{rqcc}{+ CC} & 70.67 & 98.55 & 71.56 & 94.71 & 71.37 & 96.40 \\
Qwen \textcolor{rqcc}{+ CC} & 73.67 & 98.78 & 75.03 & 88.69 & 74.74 & 93.15 \\
GLM 5.3 \textcolor{rqcc}{+ CC} & 52.83 & 93.77 & 63.34 & 92.74 & 61.10 & 93.19 \\
\bottomrule
\end{tabularx}
\end{table}

\begin{table}[!ht]
\centering
\small
\caption{RQ3 test-level pass rates (\%).}
\label{tab:test-level-rq3}
\setlength{\tabcolsep}{5pt}
\renewcommand{\arraystretch}{1.12}
\begin{tabularx}{\linewidth}{@{}l*{2}{>{\raggedleft\arraybackslash}X}@{}}
\toprule
\rowcolor{rqheader}
Condition & F2P & P2P \\
\midrule
\textcolor{rqcodex}{Joint} & 81.52 & 98.63 \\
\textcolor{rqdown}{Independent-all} & 79.90 & 97.86 \\
\bottomrule
\end{tabularx}
\end{table}

\subsection{Failure Annotation Protocol}
\label{app:failure-annotation}

We classify each unresolved run using a two-stage procedure. First, we inspect the final diffs of all target repositories. If all target repositories receive substantive code changes but the run still fails the required behavior or regression checks, the run is classified as Post-edit failure. Otherwise, we inspect the execution trajectory to distinguish the remaining two categories. If the agent fails to identify any required repository or any required change necessary to complete the task, we classify the run as Incomplete scope identification. If the agent explicitly recognizes the necessary work but does not deliver the corresponding change, we classify it as Recognized work without delivery.

Two authors independently annotated the unresolved runs according to these rules, achieving 91.2\% agreement (Cohen's $\kappa = 0.86$). Disagreements were resolved through discussion.

\section{Trajectory Evidence for Cross-Repository Mechanisms}
\label{app:trajectory-evidence}

\subsection{Cross-Repository Behavior in Joint Runs}
\label{app:codex-joint-cases}

\paragraph{Sentry SDKs.}
The task requires implementing the same strict trace-continuation policy in the Go, Python, and Ruby SDKs. Codex declares, ``The target is the Go SDK,'' and later reads Python tracing code as a reference. Only Go receives a patch, passing 14/14 \fpt{} checks; Python and Ruby pass 0/22 and 0/13. Although the SDKs have separate implementations, completing the shared request requires the agent to identify and update all three target repositories. Work on one repository can also draw on another repository's implementation, as the use of Python code while working on Go illustrates.

\paragraph{Laravel AI/MCP.}
In Laravel AI/MCP, the agent instead moves the streaming fix into the Framework context repository. AI passes 2/2 \fpt{} checks, but MCP remains unchanged at 0/3. Editing a context repository is allowed; the unresolved target behavior, rather than that location choice itself, makes the task incomplete.

\paragraph{Swagger.}
In Swagger, Codex concludes that the library's serializer ``already repeats actual JavaScript arrays correctly.'' It fixes only the UI, which passes 2/2 checks; the library remains at 0/5 because its public request-building path still mishandles the inputs.

\paragraph{Sentry CLI/Dart.}
Codex identifies the CLI parser's missing global URL option but does not identify the Dart plugin's required adaptation. It ends with an environment-variable workaround and a suggested CLI fix, leaving both repositories unmodified. Diagnosing one repository's defect does not establish the full modification scope.

\begin{figure}[p]
    \centering
    \includegraphics[width=\linewidth]{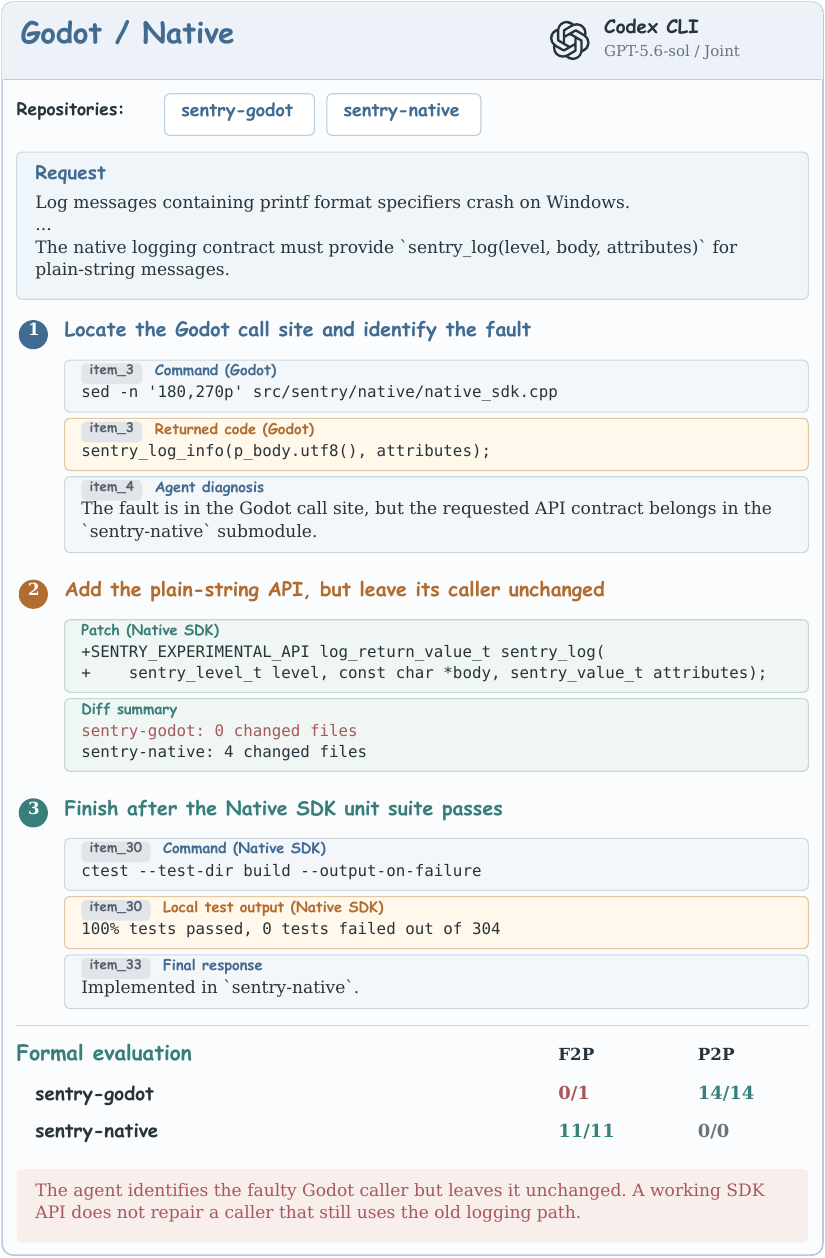}
    \caption{Godot/Native. The agent identifies the faulty Godot call site but implements and tests only the Native SDK API, leaving the caller unchanged.}
    \label{fig:godot-native-case-study}
\end{figure}

\paragraph{Godot/Native.}
In Godot/Native, Codex explicitly states that the fault is in the Godot call site, then implements only the Native API (Figure~\ref{fig:godot-native-case-study}). Native passes 11/11 \fpt{} checks while Godot remains at 0/1.

\begin{figure}[p]
    \centering
    \includegraphics[width=\linewidth]{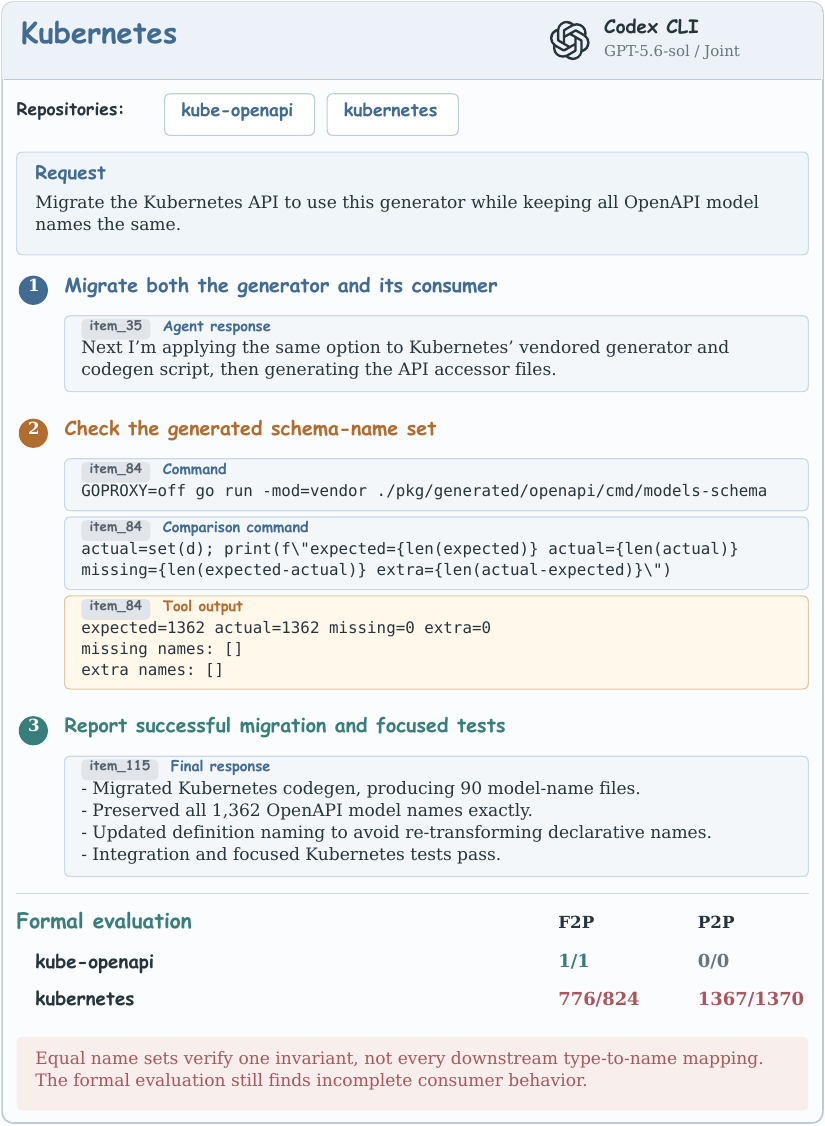}
    \caption{Kubernetes. Both generator and consumer are modified, but preserving the schema-name set does not establish complete downstream migration.}
    \label{fig:kubernetes-case-study}
\end{figure}

\paragraph{Kubernetes.}
Figure~\ref{fig:kubernetes-case-study} shows an incomplete Kubernetes consumer migration despite edits to both repositories. Preserving the generated schema names did not ensure that the downstream code used them correctly.

\paragraph{Incomplete propagation within edited repositories.}
In the Sentry Python/Ruby task, the agent edits Python's synchronous and asynchronous HTTPX integrations and Ruby's shared HTTP helper. Its targeted HTTPX command passes four tests, and the Ruby check also passes. However, Python's Celery integration still overwrites existing baggage: the evaluated message loses \texttt{custom=value}. Python passes 2/3 \fpt{} checks and Ruby 1/1. The shared requirement was propagated across repositories but not across all relevant integrations within them.

\paragraph{A shared representation can still violate the contract.}
In the Sentry/Symbolicator Java-symbolication task, the agent changes both the request producer and consumer to use an exception-name string. The required stacktrace interface instead carries a structured exception with a type and optional module. The agent's local Symbolicator suite passes 18 tests, but its integration helper leaves the new stacktrace exception at its default value, while the added remapping test supplies a string directly. Neither validates a structured exception entering the new field. Formal requests then fail with \texttt{invalid type: map, expected a string}; Symbolicator passes only 1/6 \fpt{} checks, although the other two target repositories pass theirs. This is not disagreement between the edited endpoints: both encode the same mistaken interpretation, and their local checks fail to expose it.

\paragraph{Connecting an interface without preserving its contract.}
The Sentry wizard installs the SDK's new error handler on React Router, but the handler treats a wrapped callback argument as direct React error information, losing the component stack. In Symfony's hydration migration, the agent delegates instantiation to the new DeepClone API but exposes its exceptions directly, replacing the public Symfony exceptions expected by existing callers. In both cases, the new call is present, yet the adapter fails to preserve the required input or error behavior.

\begin{figure}[p]
    \centering
    \includegraphics[width=\linewidth]{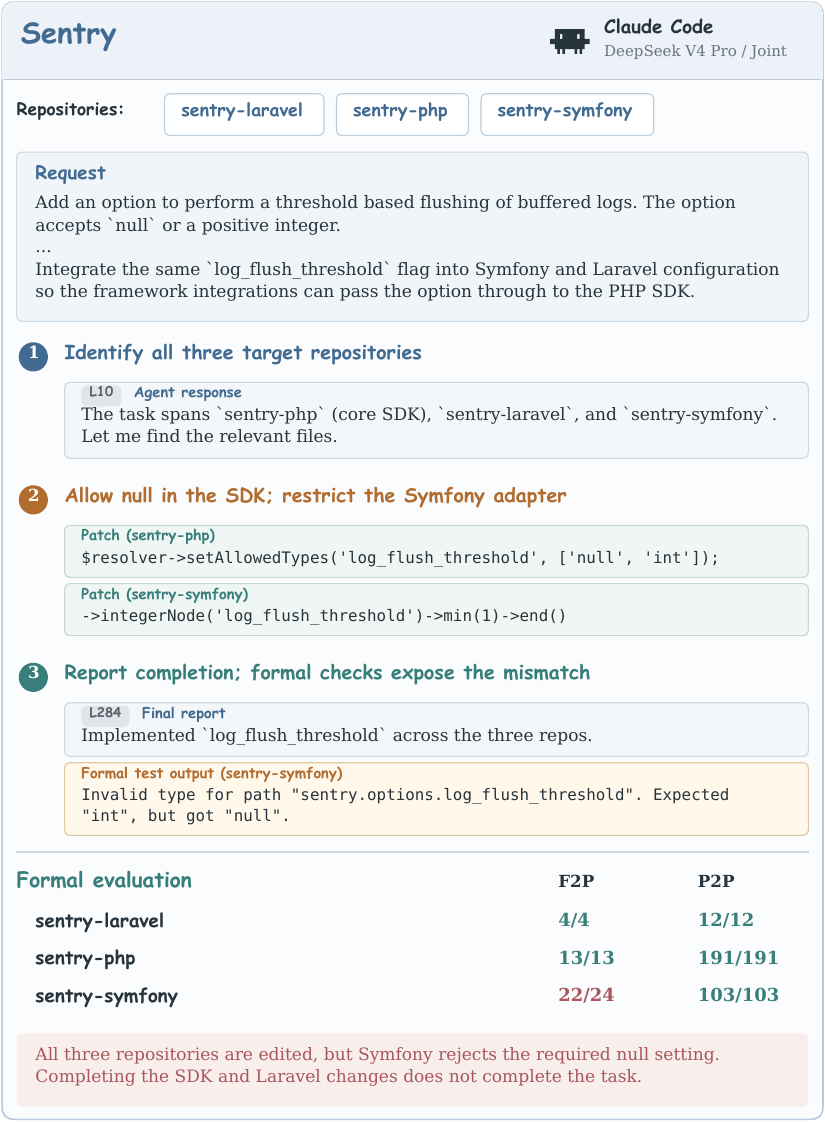}
    \caption{Sentry configuration with Claude Code--DeepSeek V4 Pro. All three repositories are edited, but Symfony rejects the required \texttt{null} setting while the PHP SDK and Laravel pass.}
    \label{fig:sentry-configuration-case-study}
\end{figure}

\paragraph{Sentry: completing two repositories while breaking the third adapter.}
The task adds \texttt{log\_flush\_threshold} to the PHP SDK and its Laravel and Symfony integrations; the option must accept \texttt{null} or a positive integer. Claude Code with DeepSeek V4 Pro identifies and modifies all three repositories (Figure~\ref{fig:sentry-configuration-case-study}). The SDK accepts \texttt{null}, but the Symfony configuration uses an integer-only node. The run reports completion, yet formal checks reject the required \texttt{null} setting in Symfony. PHP and Laravel pass 13/13 and 4/4 \fpt{} checks, respectively, while Symfony passes 22/24. This failure concerns the correctness of the remaining adaptation, not an overlooked repository.

\begin{figure}[p]
    \centering
    \includegraphics[width=\linewidth]{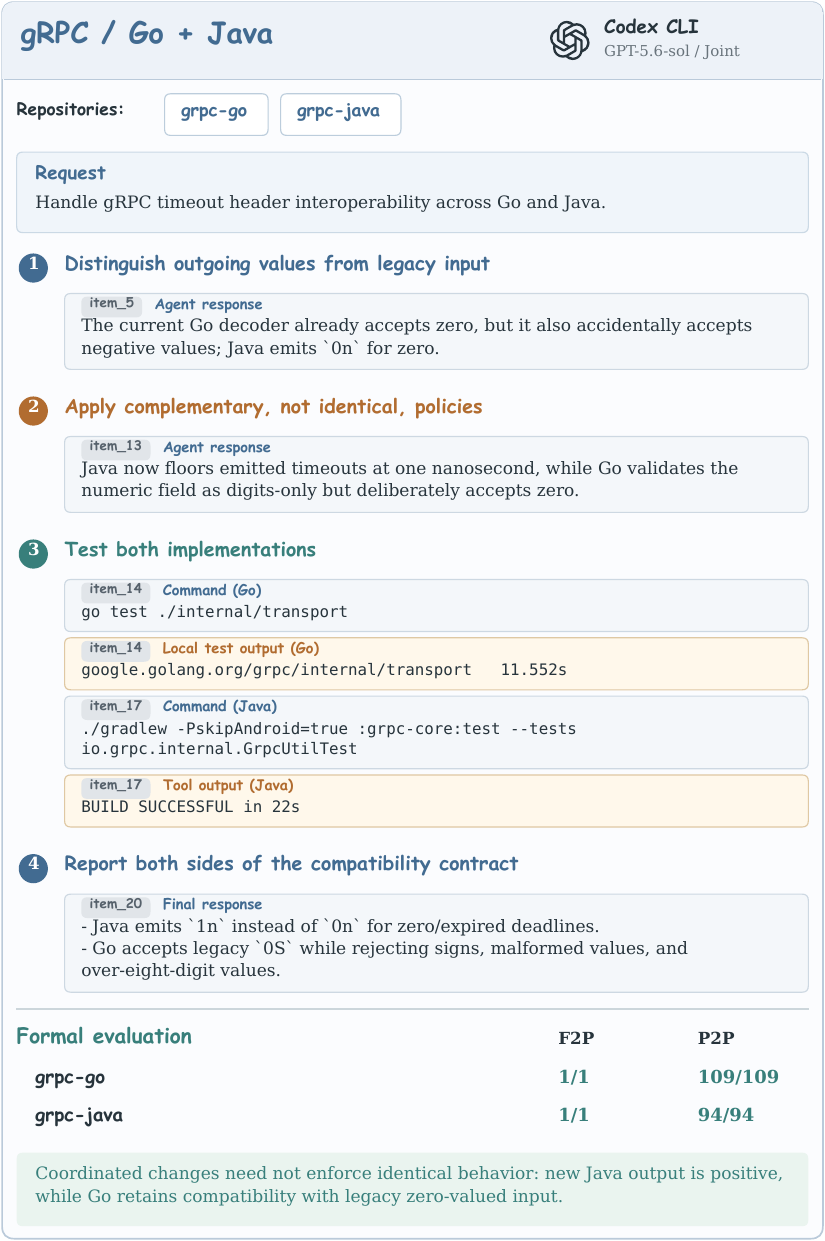}
    \caption{gRPC. Compatibility requires complementary policies: Java emits positive timeouts, while Go accepts legacy zero-valued input. Both repositories pass formal evaluation.}
    \label{fig:grpc-case-study}
\end{figure}

\paragraph{Findings and implications.}
The cases distinguish identifying necessary changes, covering relevant execution paths, and preserving the contracts between repositories. Godot/Native leaves the diagnosed caller unchanged; the Python/Ruby task leaves an integration unfinished inside an edited repository; and Sentry/Symbolicator adopts a consistent but incorrect request representation. Conversely, gRPC succeeds by applying complementary, rather than identical, rules at the two endpoints (Figure~\ref{fig:grpc-case-study}). Neither repository coverage nor agreement between edited components alone establishes coordinated completion.

\subsection{Joint versus Independent Codex Runs}
\label{app:scope-case-comparison}

These three tasks illustrate omitted-work recovery and two uses of cross-repository context under the paired setup in Appendix~\ref{app:joint-independent}.

\begin{figure}[p]
    \centering
    \includegraphics[width=\linewidth]{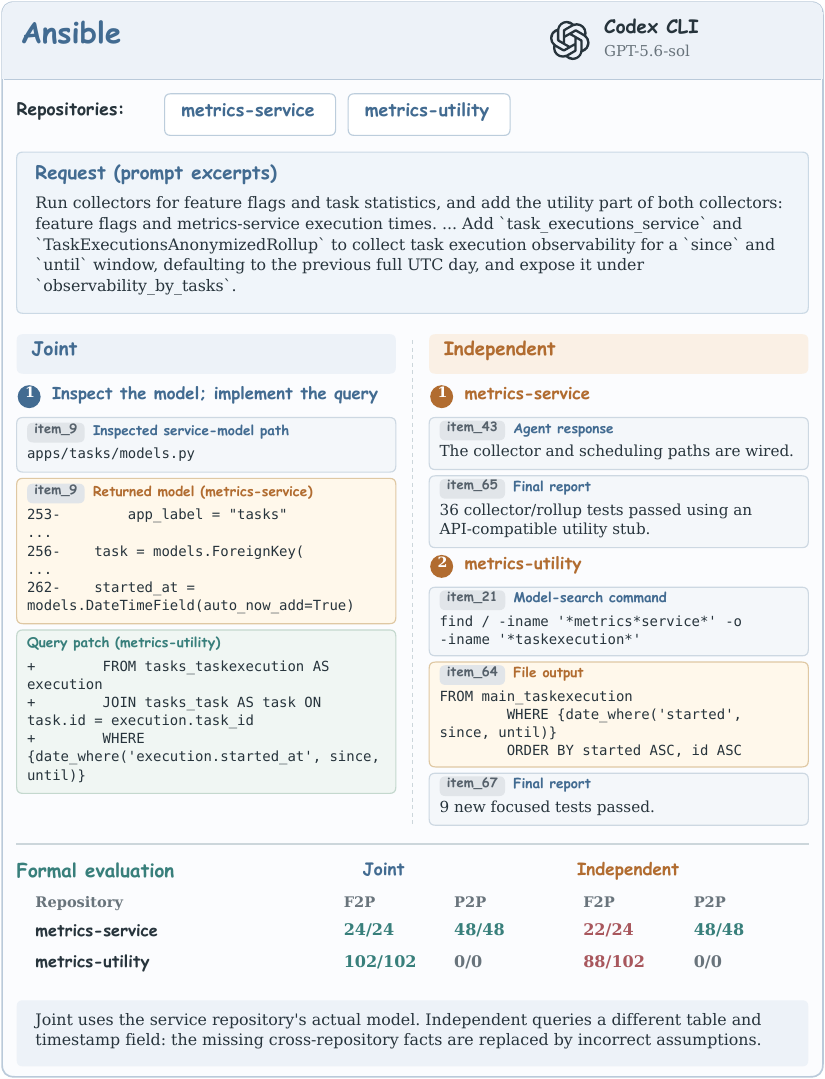}
    \caption{Ansible. The Joint run reads the service repository's model and implements a compatible utility query. The independent utility run submits a query inconsistent with that model.}
    \label{fig:ansible-case-study}
\end{figure}

\paragraph{Ansible: using the other repository to guide implementation.}
The task requires metrics-utility to collect execution information from metrics-service. The Joint run reads the service's task-execution model before implementing the utility query. Its query uses the service's actual task and execution tables and the \texttt{started\_at} field. The independent utility run searches for the service model but submits a query using a different table and a \texttt{started} field instead. Figure~\ref{fig:ansible-case-study} contrasts the inspected model with the submitted queries. This illustrates how another repository supplies concrete implementation facts, rather than only additional work to complete.

\paragraph{Findings and implications.}
Separate assignments recover omitted SDK work in Sentry. In Ansible, the related repository supplies facts needed to implement a compatible query; in Symfony, it provides a behavioral reference that helps reveal an implementation error. The cases distinguish completing overlooked work from using cross-repository information to implement and check that work correctly.

\begin{figure}[p]
    \centering
    \includegraphics[width=\linewidth]{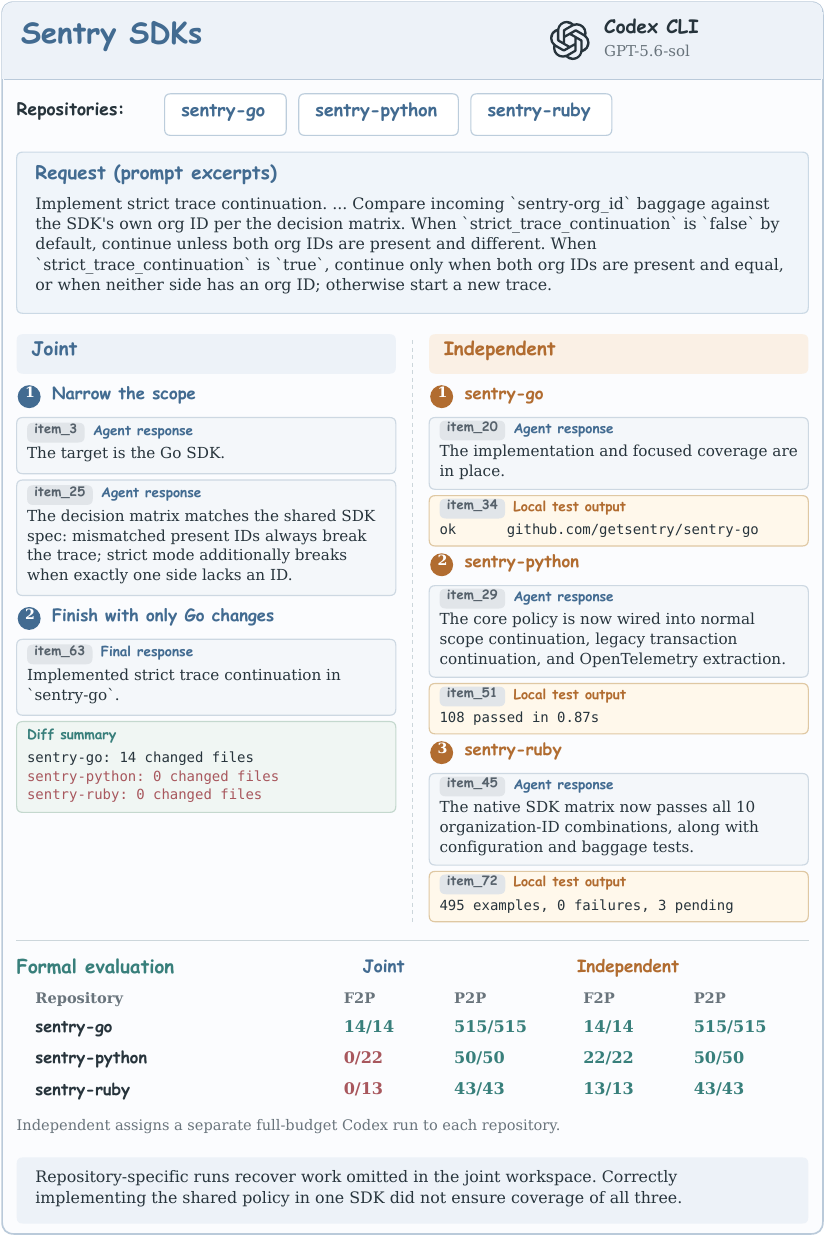}
    \caption{Sentry. The joint run implements a shared policy only in Go. Independent runs complete all three SDKs, recovering the previously omitted Python and Ruby work.}
    \label{fig:sentry-case-study}
\end{figure}

\begin{figure}[p]
    \centering
    \includegraphics[width=\linewidth]{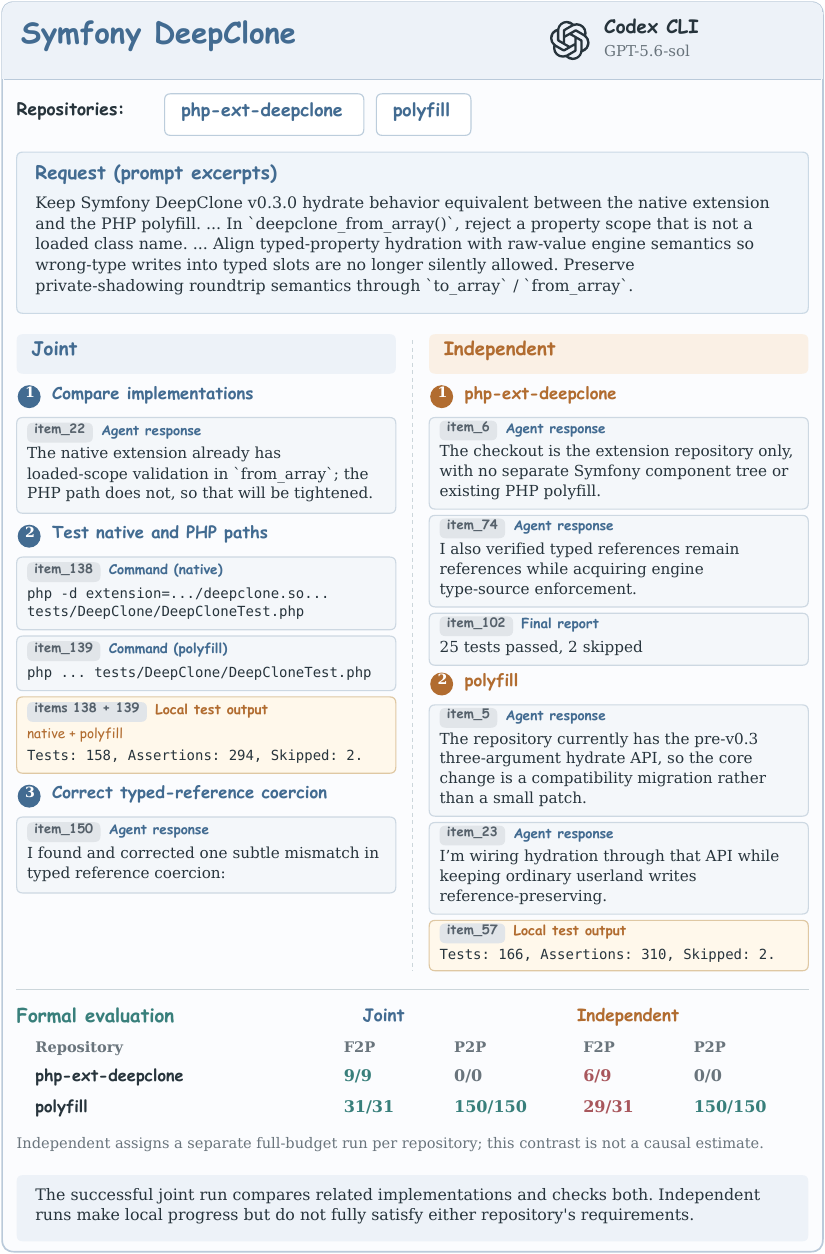}
    \caption{Symfony. The successful joint run compares native and PHP behavior and tests both implementations. Independent runs pass local checks but leave required behavior incomplete in both repositories.}
    \label{fig:symfony-case-study}
\end{figure}

\subsection{Codex CLI versus Claude Code with GPT-5.6-sol}
\label{app:scaffold-comparison}

The Rails/Propshaft pair holds the model and prompt fixed. It illustrates a difference in scope checking between two runs, rather than isolating a particular framework component.

\begin{figure}[p]
    \centering
    \includegraphics[width=\linewidth]{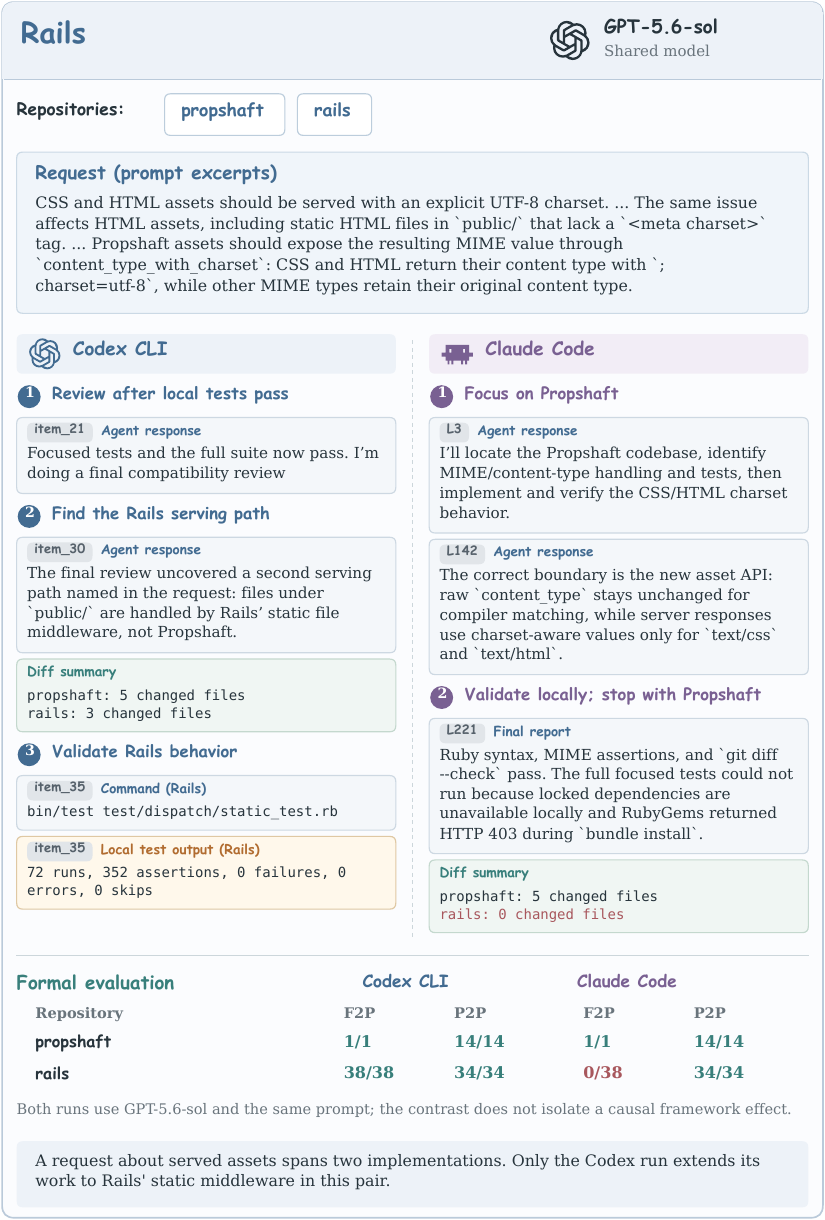}
    \caption{Rails/Propshaft. Codex revisits the request after Propshaft tests pass and extends the change to Rails' static-file middleware. Claude Code with the same GPT model modifies only Propshaft.}
    \label{fig:rails-case-study}
\end{figure}

\subsection{Model Comparisons within Claude Code}
\label{app:model-comparison}

Each comparison holds Claude Code and the task prompt fixed while varying the LLM. Fleet illustrates the Delivery failure category in RQ1: recognized work is not implemented. Rust/Cargo, PyPA, and TensorDict/TorchRL illustrate Post-edit failures: the target repositories are modified, but the changes do not jointly satisfy the request. These selected runs illustrate the failure mechanisms rather than establish model-wide rankings.

\begin{figure}[p]
    \centering
    \includegraphics[width=\linewidth]{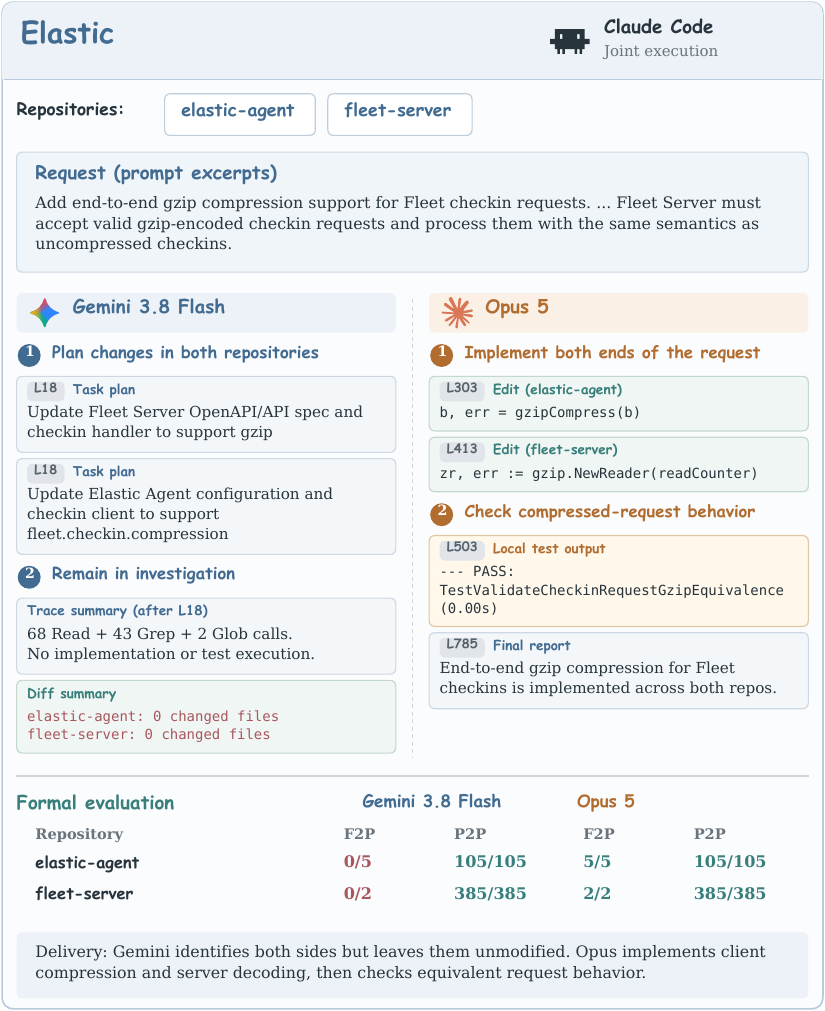}
    \caption{Elastic with Claude Code. Gemini plans changes in both repositories but remains in investigation. Opus implements client compression and server decoding, checks equivalent request behavior, and passes both repositories.}
    \label{fig:fleet-case-study}
\end{figure}

\paragraph{Fleet: carrying a plan through to implementation.}
The task requires Elastic Agent to compress checkin requests and Fleet Server to accept them. Gemini plans both changes but then makes 113 read or search calls without implementing either side. Opus implements client compression and server decoding, then checks that compressed and uncompressed requests behave equivalently (Figure~\ref{fig:fleet-case-study}). Gemini passes 0/5 and 0/2 \fpt{} checks in the two repositories; Opus passes 5/5 and 2/2. The contrast is between recognizing the required work and carrying it out, not between identifying different target repositories.

\begin{figure}[p]
    \centering
    \includegraphics[width=\linewidth]{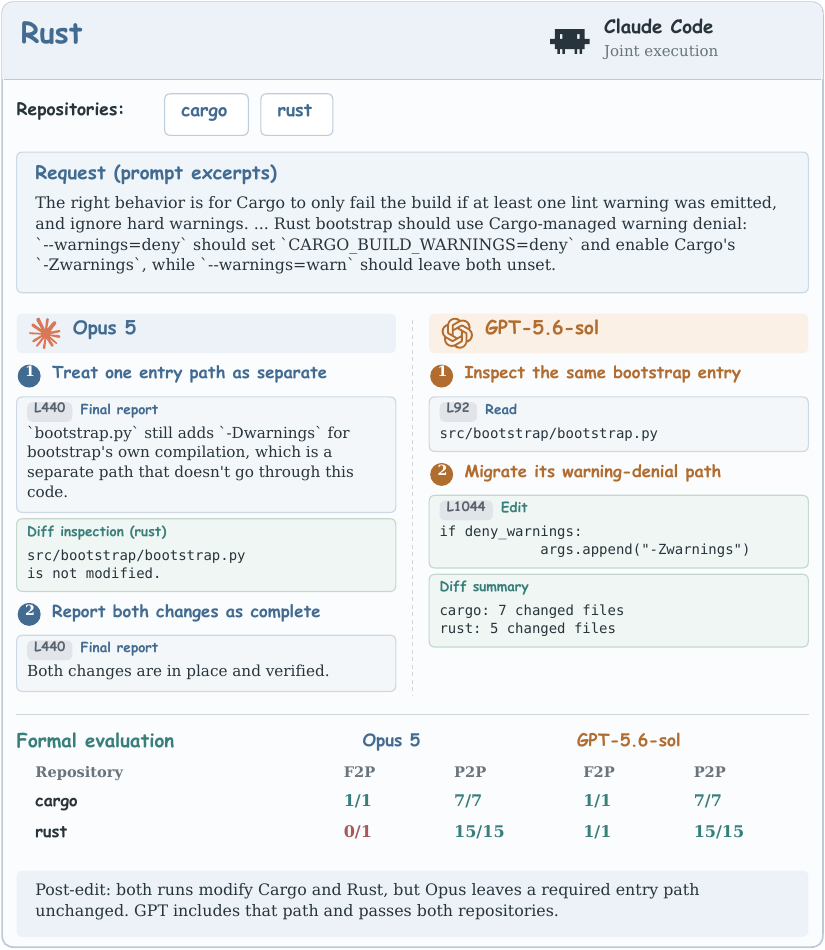}
    \caption{Rust with Claude Code. Both runs complete the Cargo change. Opus leaves the Python bootstrap entry unchanged, while GPT includes it in the migration and passes both repositories.}
    \label{fig:rust-case-study}
\end{figure}

\paragraph{Rust/Cargo: covering the consumer's entry points.}
The task requires Rust bootstrap to adopt Cargo-managed warning denial. Opus and GPT both modify Cargo and Rust, but Opus treats \texttt{bootstrap.py} as a separate path and leaves it unchanged, whereas GPT includes it in the migration (Figure~\ref{fig:rust-case-study}). Opus passes Cargo's evaluation but fails Rust's \fpt{} check; GPT passes both repositories. DeepSeek also migrates this entry path and passes both. Because the failing Opus run edits both target repositories, this is incomplete implementation within an edited repository, not failure to identify a target repository.

\begin{figure}[p]
    \centering
    \includegraphics[width=\linewidth]{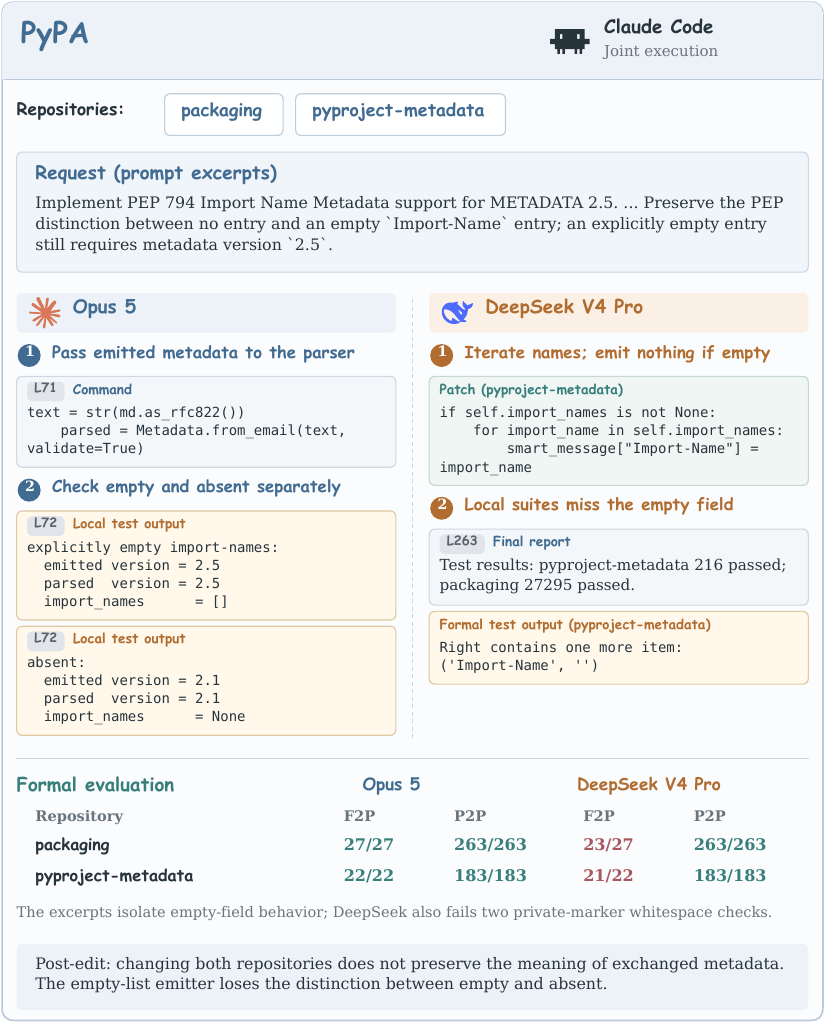}
    \caption{PyPA. Opus checks that metadata emitted by pyproject-metadata retains its meaning when parsed by packaging. DeepSeek changes both repositories and reports passing local suites, but its emitter drops explicitly empty import-name fields.}
    \label{fig:pypa-case-study}
\end{figure}

\paragraph{PyPA: checking exchanged metadata.}
The task distinguishes an absent \texttt{Import-Name} field from an explicitly empty one. Opus feeds the producer's output into the parser and checks both cases. DeepSeek modifies both repositories and reports passing local suites, but its emitter drops the empty field (Figure~\ref{fig:pypa-case-study}). Opus passes both repositories; DeepSeek leaves both \fpt{}-incomplete.

\begin{figure}[p]
    \centering
    \includegraphics[width=\linewidth]{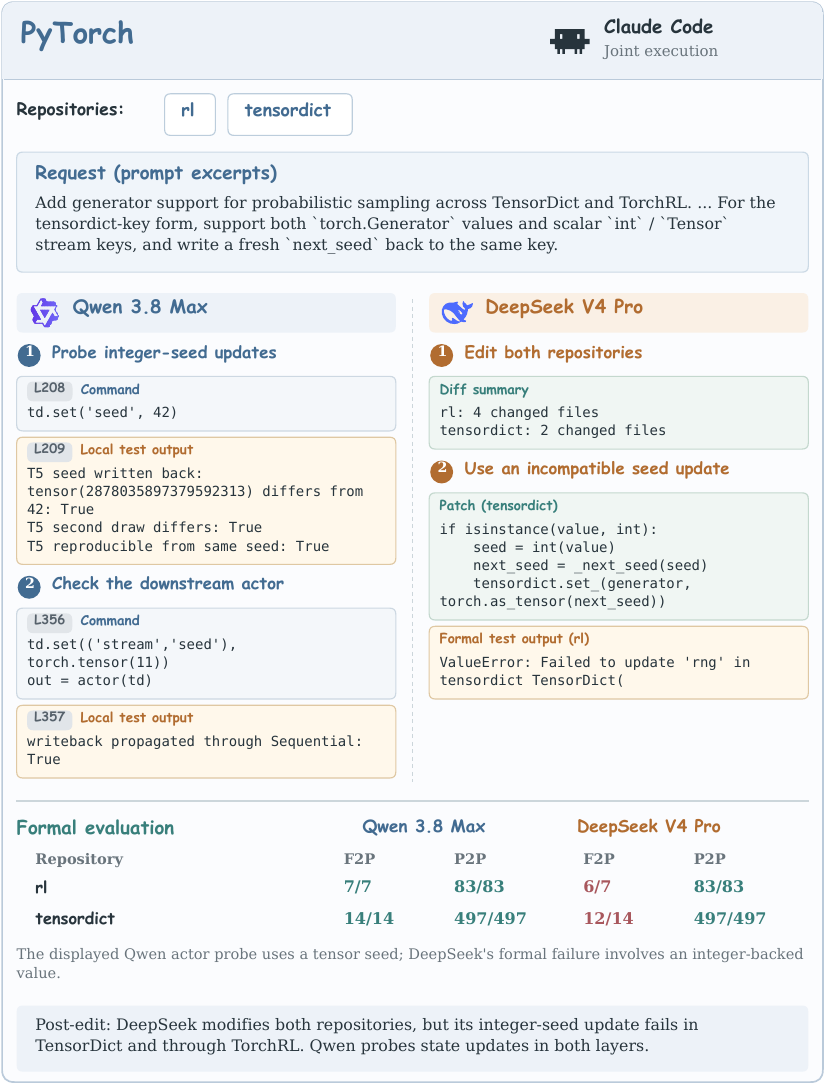}
    \caption{TensorDict/TorchRL. Qwen probes seed updates in the library and through its downstream actor. DeepSeek also modifies both repositories, but an integer-seed write-back error affects both layers, illustrating how a shared-state defect propagates across repository boundaries.}
    \label{fig:pytorch-case-study}
\end{figure}

\paragraph{TensorDict/TorchRL: checking downstream state updates.}
The task requires sampling to write a fresh seed back to the specified TensorDict key. Qwen probes seed updates in TensorDict and through a TorchRL actor. DeepSeek also edits both repositories, but its integer-seed write-back raises an error in both layers (Figure~\ref{fig:pytorch-case-study}). Qwen passes both repositories, while DeepSeek leaves both \fpt{}-incomplete.

\section{Benchmark Scope and Limitations}
\label{app:limitations}

\benchmark{} tasks are mined from real-world cross-repository changes and involve two or three target repositories. They cover diverse coordination patterns, including shared feature implementation across language SDKs, adaptations between libraries and downstream integrations, and interface changes between producers and consumers. Tasks are evaluated in Linux environments. The results do not establish how agents perform on tasks involving more repositories or requiring other platforms.

\end{document}